\documentclass{emulateapj}

\usepackage{natbib}
\usepackage{amsmath}

\shorttitle{Star Formation in II~Zw~40}
\shortauthors{Kepley et al.}

\newcommand{\ha}{\ensuremath{{\rm H}\alpha}}

\newcommand{\Msun}{\ensuremath{\rm{M}_\odot}}

\newcommand{\uJybeam}{\ensuremath{\rm{\mu Jy \ beam^{-1}} \ }}
\newcommand{\mJybeam}{\ensuremath{\rm{mJy \ beam^{-1}} \ }}

\newcommand{\Av}{\ensuremath{{\rm A_V}}}

\newcommand{\hii}{{\rm H\,}{{\sc ii}}}

\begin{document}

\renewcommand{\floatpagefraction}{0.90}


\title{High Resolution Radio and Optical Observations of the \\
  Central Starburst in the Low-Metallicity Dwarf Galaxy II~Zw~40}

\author{Amanda A. Kepley\altaffilmark{1}}
 
\affil{Department of Astronomy, University of Virginia, P.O. Box
  400325, Charlottesville, VA 22904-4325, USA} 

\email{akepley@nrao.edu}

\author{Amy E. Reines\altaffilmark{2}}

\affil{Department of Astronomy, University of Virginia, P.O. Box
  400325, Charlottesville, VA 22904-4325, USA; and National Radio Astronomy
  Observatory, 520 Edgemont Road, Charlottesville, VA 22903, USA}
\email{areines@nrao.edu}

\author{Kelsey E. Johnson, Lisa May Walker}

\affil{Department of Astronomy, University of Virginia, P.O. Box
  400325, Charlottesville, VA 22904-4325, USA}
\email{kej7a@virginia.edu, lisamay@virginia.edu}

\altaffiltext{1}{currently at the National Radio Astronomy
  Observatory, P.O. Box 2, Green Bank, WV 24944, USA}
\altaffiltext{2}{Einstein Fellow}

\begin{abstract}

  The extent to which star formation varies in galaxies with low
  masses, low metallicities, and high star formation rate surface
  densities is not well-constrained. To gain insight into star
  formation under these physical conditions, this paper estimates the
  ionizing photon fluxes, masses, and ages for young massive clusters
  in the central region of II~Zw~40 -- the prototypical
  low-metallicity dwarf starburst galaxy -- from radio continuum and
  optical observations. Discrete, cluster-sized sources only account
  for half the total radio continuum emission; the remainder is
  diffuse. The young ($\lesssim 5$~Myr) central burst has a star
  formation rate surface density that significantly exceeds that of
  the Milky Way. Three of the 13 sources have ionizing
  photon fluxes (and thus masses) greater than R136 in 30 Doradus.
  Although isolating the effects of galaxy mass and metallicity is
  difficult, the \hii\ region luminosity function and the internal
  extinction in the center of II~Zw~40 appear to be primarily driven
  by a merger-related starburst. The relatively flat \hii\ region
  luminosity function may be the result of an increase in ISM pressure
  during the merger and the internal extinction is similar to that
  generated by the clumpy and porous dust in other starburst galaxies.

\end{abstract}



\section{Introduction} \label{sec:introduction}

Dwarf starburst galaxies have the potential to provide vital
information about how star formation proceeds when subject to physical
conditions not present in the Milky Way. These galaxies typically have
lower masses, lower metallicities, and much higher star formation rate
surface densities than spiral galaxies, which may change what physical
processes are important for star formation and the properties of the
stars and clusters in the galaxy.

The effects of the unique environment for star formation found in
dwarf starburst galaxies can manifest in several ways. First, the
clusters in these galaxies may be larger than the clusters in normal
galaxies due to either lack of large-scale shear from differential
rotation \citep{2010ApJ...724.1503W}, higher gas pressures
\citep{1997ApJ...480..235E}, or high gas densities decreasing the
effectiveness of radiation pressure for fragmenting molecular clouds
\citep{2010ApJ...713.1120K}. Second, the molecular gas tracers in
these galaxies may have different properties than the molecular gas
tracers in more normal galaxies: there are fewer metals overall, less
dust to serve as a formation site for molecular hydrogen, and the
harder radiation fields may destroy more dust
\citep{2003A&A...407..159G,2005A&A...434..867G,2006A&A...446..877M}
and molecules \citep[e.g.][]{2008ApJ...686..948B}. Finally, star
formation may be regulated by different mechanisms in dwarf starburst
galaxies including the radiation pressure and/or \hii\ region
expansion \citep{2010ApJ...709..191M} and the injection of turbulence
by supernovae \citep{2011ApJ...731...41O}.

The prototypical dwarf starburst galaxy II~Zw~40 provides an important
test case for models of star formation in low-mass, low-metallicity,
and high star formation rate surface density environments. This
galaxy, along with I Zw 18, originally defined the blue compact dwarf
galaxy class, which consists primarily of low mass, low metallicity,
starbursting galaxies
(\citealp{1970ApJ...162L.155S,1972ApJ...173...25S}; see
\citet{2003ApJS..147...29G} for an updated definition). The total
dynamic mass of II~Zw~40 ($6\times10^9 M_\odot$;
\citealp{1988MNRAS.231P..63B}) is only 4\% of the total dynamic mass
of the Milky Way. Furthermore, II~Zw~40 has a metallicity of only 1/5
Z$_\odot$ \citep[$≈ƒ12 + \log({\rm O/H}) =
8.09$;][]{2000ApJ...531..776G}.  Finally, the star formation rate
surface density for this entire dwarf galaxy is $0.4 \ {\rm M_\odot \,
  yr^{-2} \, kpc^{-2}}$, which is comparable to other starburst
galaxies \citep[cf. Figure 15 in][]{2008AJ....136.2846B}.

Even in low-metallicity environments, young star-forming regions can
still be subject to significant internal extinction, making them
difficult to study in the optical and infrared
\citep{1999ApJ...516..783T,2002AJ....124.1995P,2003A&A...407..159G,2005A&A...434..867G}.
In the radio, however, free-free emission from the ionized gas
surrounding young massive stars is minimally affected by extinction by
dust. Consequently, free-free emission provides us with an important
tool for inferring the properties of the young massive clusters that
may be partially or totally obscured at optical wavelengths. It is not
perfect tool though: the number of ionizing photons produced by the
obscured, young, massive stars may be underestimated because these
photons are absorbed by dust or escape from the region (see the
discussion in Section~\ref{sec:phys-prop-radio}).

The goal of this paper is to quantify the properties of star formation
in the low-metallicity, low-mass, and high star formation rate surface
density environment found in II Zw 40.  We present measurements of the
young massive cluster population in II~Zw~40 using new 6.2\,cm,
3.5\,cm, and 1.3\,cm radio continuum images from the National Radio
Astronomy Observatory (NRAO) Very Large Array (VLA)\footnote{The
  National Radio Astronomy Observatory is a facility of the National
  Science Foundation operated under cooperative agreement by
  Associated Universities, Inc.} (\S~\ref{sec:vla-data}) and
previously unpublished high-resolution optical F555W ($\sim$V-band),
F658N (\ha), and F814W ($\sim$I-band) images from the Advanced Camera
for Surveys (ACS) on the {\it Hubble Space Telescope
  (HST)}\footnote{Based on observations made with the NASA/ESA Hubble
  Space Telescope, obtained from the Data Archive at the Space
  Telescope Science Institute, which is operated by the Association of
  Universities for Research in Astronomy, Inc., under NASA contract
  NAS 5-26555. These observations are associated with program \#6739.}
(\S~\ref{sec:hst-acs}). We describe the properties of the young
massive clusters as seen in the radio continuum
(\S~\ref{sec:prop-radio-cont}) and the optical
(\S~\ref{sec:prop-optic-clust}). Then we discuss the implications of
our results including the cluster luminosity function as traced by the
\hii\ region luminosity function (\S~\ref{sec:hii-regi-lumin}) and the
internal dust extinction (\S~\ref{sec:dust-properties}). Finally, we
summarize our results and present our conclusions and their
implications (\S~\ref{sec:discussion}).

This paper assumes a distance to II~Zw~40 of 10~Mpc
\citep{1988cng..book.....T}. At this distance, 0.1\arcsec\ corresponds
to 4.8~pc.

\section{Data} \label{sec:data}

To trace the obscured young massive star formation in II~Zw~40, we
obtained high resolution radio continuum observations at 6.2\,cm,
3.5\,cm, and 1.3\,cm with the VLA. The array configurations were
chosen to produce relatively matched {\em u-v} coverage at each
frequency. These new observations have the sensitivity and resolution
necessary to probe the physical scales and flux densities of embedded
star forming regions; they are 2.4 to 7 times more sensitive in
surface brightness, 4 to 20 times deeper in point source sensitivity,
and 2 to 3 times higher resolution than the previous highest
resolution radio continuum survey
\citep{2002AJ....124.2516B}.\footnote{We included some of the archival
  data from \citet{2002AJ....124.2516B} to increase sensitivity and
  avoid unnecessarily duplicating observations.}
Table~\ref{tab:vla_obs_summary} summarizes the radio continuum data
used in this paper.  The calibration and imaging of these data are
detailed in \S~\ref{sec:vla-data}

To obtain a more complete picture of star formation in II Zw 40, this
paper also includes high resolution optical F555W (similar to V-band),
F658N (\ha), and F814W (similar to I-band) observations from the {\em
  HST}. We use these data to parameterize the unobscured young massive
clusters in II~Zw~40 -- in particular, the two bright optically visible
SSCs. The optical data also measure the \ha\ emission from the ionized
gas. More detailed information on the {\em HST} data is provided in
\S~\ref{sec:hst-acs}.

\begin{figure*}
\centering
\includegraphics{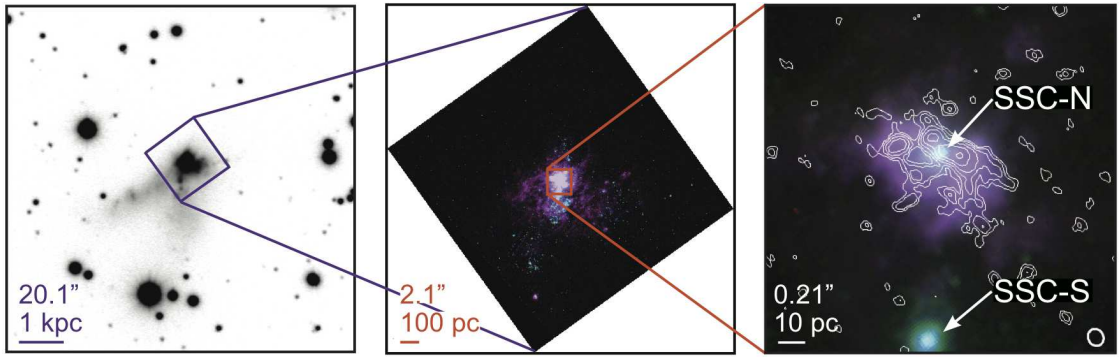}
\caption{{\em Left:} R band image of II~Zw~40 from
  \citet{2003ApJS..147...29G}. The two linear features extending to
  the east and south are suggestive of tidal tails.  {\em Middle:}
  {\it HST} ACS image of II~Zw~40 from archival observations by R.\
  Chandar. Blue represents the F555W filter (V-band), green represents
  the F814W filter (I-band), and red represents the F658N filter
  (\ha). Note the extended diffuse \ha\ component throughout the
  central region of the galaxy. {\em Right:} Contours from our 1.3\,cm
  robust~=~0 image overlaid on the central region of the ACS image
  with the same color scheme as in the middle panel. The contours
  start at the 3$\sigma$ noise level for the 1.3\,cm robust~=~0 image
  and increase by factors of $2^{n/2}$, where $n = 1,2,3,...$. The
  beam for the 1.3\,cm robust~=~0 image is given in the lower right
  hand corner of the image. The two optical sources (SSC-N and SSC-S)
  are discussed in \S~\ref{sec:prop-optic-clust}.}
\label{fig:overview}
\end{figure*}

Figure~\ref{fig:overview} gives an overview of the observations
presented here, illustrating the overall morphology and size of
II~Zw~40 and the central few hundred parsec region of II~Zw~40 probed
by our observations.

\subsection{VLA Data} \label{sec:vla-data}

The details for the VLA observations are given in
Table~\ref{tab:vla_obs_summary}. The use of the Pie Town link, a Very
Long Baseline Array (VLBA) antenna located approximately 52 km away
from the main VLA site, doubles the resolution of the VLA A
configuration \citep{VLAtestmemo217}. For all the observations, the
correlator was set up to provide two 50~MHz IFs.  These IFs were tuned
to 4.8851 and 4.8351~GHz, 8.4351~GHz and 8.4851~GHz, and 22.4851 and
22.4351 GHz for the 6.2\,cm, 3.5\,cm, and 1.3\,cm observations,
respectively. For the 1.3\,cm observations, reference pointing was
done every hour to minimize pointing errors and fast switching mode
was used to minimize atmospheric phase variations. We have removed the
4 EVLA antennas from the August 2006 data set to avoid introducing
phase closure errors in the crossed (EVLA-VLA) baselines because of
mismatches in the bandpasses of the VLA and EVLA receivers.

The 6.2\,cm and 3.5\,cm data were reduced using the standard AIPS
reduction procedures detailed in Appendix A of the AIPS Cookbook
\citep{ch2:aipscookbook}. Here we provide a sketch of the
calibration. The data were read in with FILLM and the flux density of
the primary calibrator was established using SETJY. Then the amplitude
and phases of the primary and secondary calibrator were determined
using CALIB; a model was used for the primary calibrator and the
secondary calibrator was assumed to be a point source. The flux
density of the secondary calibrators were determined using
GETJY. Finally, the calibration solutions were applied using CLCAL.

The 1.3\,cm data were reduced following the procedure in Appendix D in
the AIPS Cookbook \citep{ch2:aipscookbook}.  First, the data were read
in and had atmospheric corrections applied based on current and
seasonal weather data using the task FILLM. The positions of the
antennas were corrected using the task VLANT. Then, we set the flux
density of the primary calibrator using SETJY. We calibrated the
phases of the primary and secondary calibrators with CALIB using a
solution interval of 30s, again using a model for the primary
calibrator and assuming the secondary calibrator is a point
source. For the August 2006 observations, the {\it u-v} range for the
solutions for the secondary calibrator was restricted to baselines
less than 300 k$\lambda$. The other data sets were taken during better
conditions and the {\it u-v} range of the solutions did not need to be
restricted to produce a good calibration.  We applied these phase
calibrations using CLCAL. Then we calibrated the amplitudes and phases
of the primary and secondary calibrators with CALIB using the scan
length as the solution interval. Again, for the August 2006 data set
we restricted the solution for the secondary calibrators to baselines
to less than 300 k$\lambda$. We bootstrapped the flux density of the
secondary calibrators from the flux density of the primary calibrator
using GETJY. Finally, we applied the antenna-based amplitude and phase
calibration using CLCAL.


The process of creating images from the visibilities measured by an
interferometer depends on the scientific goals of the project and the
properties of the data. Our two goals for this data -- to measure the
radio continuum spectrum and to identify individual radio continuum
sources -- required us to produce two different sets of images.

\begin{figure*}
\centering
\includegraphics[width=\textwidth]{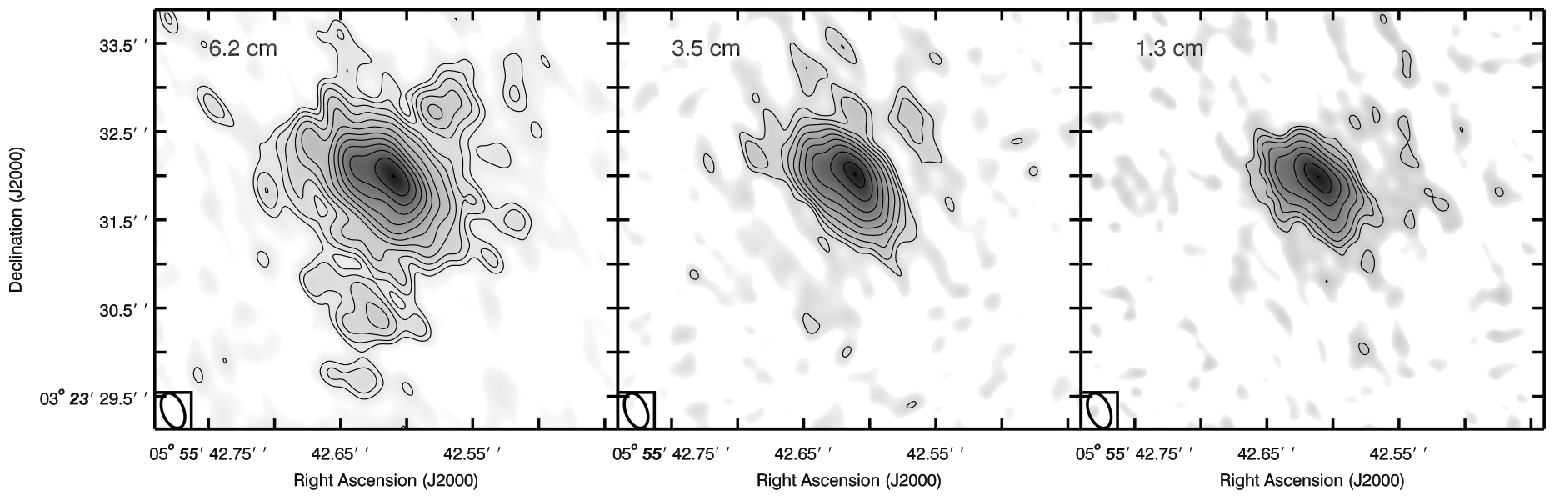}
\caption{Matched {\em u-v} coverage images of our data at 6.2\,cm
  ({\em left}), 3.5\,cm ({\em middle}), and 1.3\,cm ({\em right}).
  The contours start at the 3$\sigma$ noise level for each image and
  increase by factors of $2^{n/2}$, where $n = 1,2,3,...$. The beam is
  shown in the lower left hand corner of each panel.}
\label{fig:cxk_matched}
\end{figure*}

To measure the global radio continuum spectrum, we matched the spatial
resolutions and {\it u-v} coverage at different wavelengths to reduce
systematic effects from differences in spatial sampling. The matched
{\em u-v} coverage 6.2\,cm, 3.5\,cm, and 1.3\,cm images are shown on
the same spatial scale in Figure~\ref{fig:cxk_matched} and image
properties are given in Table~\ref{tab:image_summary}. We note that
cleaning the images does not change the derived total flux. The total
flux in an image with a zero-spacing flux value of 18 mJy (the
\citealt{1991A&A...246..323K} value for the total 22~GHz flux) is
within 0.2\% of the total flux in an image with zero-spacing flux
value of 0 mJy.

\begin{figure*}
\centering
\includegraphics{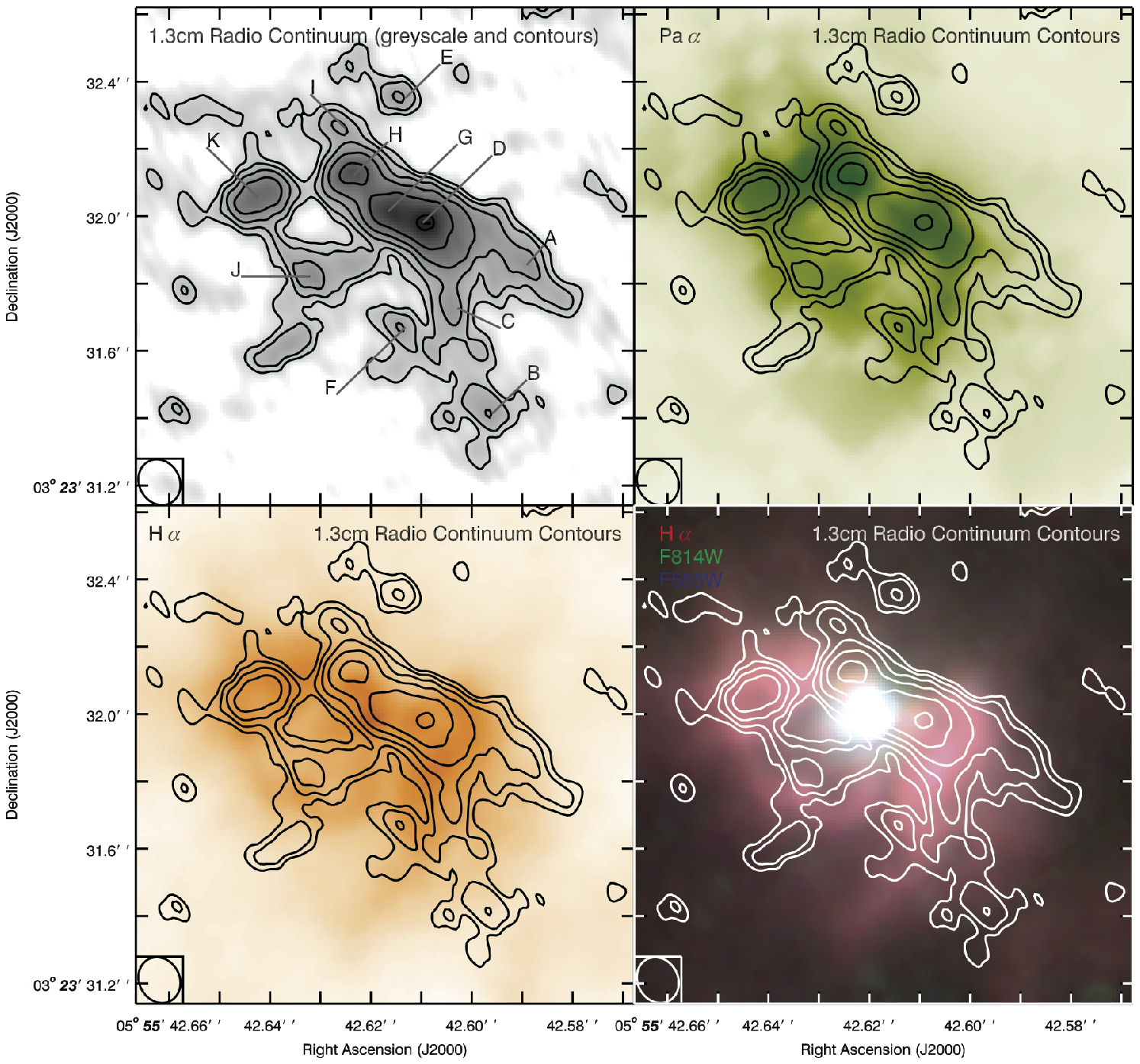}
\caption{{\em Top Left:} Zoomed in version of 1.3\,cm robust~=~0
  image.  The contours start at the 3$\sigma$ noise level and increase
  by factors of $2^{n/2}$, where $n = 1,2,3,...$. The beam is shown in
  the lower left hand corner. The same contours and beam are shown in
  the other three panels. The letters indicate the sources identified
  in Section~\ref{sec:radio-source-prop} and are ordered by increasing
  RA. {\em Top Right:} Continuum subtracted NICMOS Paschen $\alpha$
  image of II~Zw~40. {\em Bottom Left:} The F658N ACS image, which
  corresponds to H$\alpha$, of II~Zw~40. The F658N image is not
  continuum subtracted.  {\em Bottom Right:} Three color optical image
  of II~Zw~40. In this image, blue is used for the ACS F555W filter
  image (roughly V-band), green is used for the F814W filter image
  (roughly I-band), and red is used for the ACS F658N image (\ha).  }
\label{fig:kband_hst_4panel}
\end{figure*}

The spatial resolution of the matched beam images is 12 to 20 pc: too
large to probe the physical size scales of natal clusters. To identify
and measure the properties of individual radio continuum sources, we
produced a high resolution, robust=0 image of the 1.3\,cm continuum
data with a physical resolution of 6.7 by 5.8 pc: two to three times
higher than the resolution of the matched images and comparable to the
size scales of individual clusters.  The 1.3\,cm robust~=~0 image is
shown in Figure~\ref{fig:kband_hst_4panel} and the image properties
are given in Table~\ref{tab:image_summary}.

\subsection{{\it HST} ACS data} \label{sec:hst-acs}

{\it HST} observations of II~Zw~40 were obtained with the ACS High
Resolution Channel (HRC) on 2003 November 26 ({\it HST} proposal 6739,
PI: R.\ Chandar).  Images were taken with the F555W, F658N, and
F814W filters.  We retrieved the pipeline-produced calibrated and
drizzled images from the archive.  The HRC plate scale is $\sim
0\farcs025$ pixel$^{-1}$ and the images have a resolution of $\sim
0\farcs075$.

The F555W and F814W filters are broad and can be contaminated by
nebular emission from young ($\lesssim 5$ Myr) clusters
\citep{2010ApJ...708...26R}.  In addition, at the redshift of
II~Zw~40, the F658N narrow-band filter contains both the H$\alpha$ and
[NII]$\lambda$6584 emission lines.  However, spectroscopic
observations of the brightest part of the galaxy show that the
[NII]$\lambda$6584 flux density is only $\sim 2\%$ of the H$\alpha$
flux density \citep{2000ApJ...531..776G}. Therefore, we expect the
contamination from this line to be negligible.

The nominal astrometry of the {\it HST}/ACS images is only accurate to
within $\sim 1\arcsec$ while the VLA astrometry is good to within
$\sim 0\farcs1$ and is considered absolute by comparison.  We
therefore register the optical {\it HST} images to the high resolution
VLA 1.3\,cm image by bootstrapping from a continuum-subtracted {\it
  HST} NICMOS Paschen $\alpha$ image which traces the same ionized gas
as the thermal radio emission (modulo extinction).  The F658N ACS
image containing both H$\alpha$ emission and stellar continuum was
then registered to the Paschen $\alpha$ image.  The F555W and F814W
ACS images are aligned with the registered F658N image. We estimate
that the final relative astrometry is accurate to $\lesssim0.1\arcsec$
based on a comparison of the Paschen $\alpha$ data and the radio continuum
data.

Figure~\ref{fig:kband_hst_4panel} shows the distribution of the
re-registered {\it HST} ACS F658N, F555W, and F814W and the NICMOS
Paschen $\alpha$ compared to our highest resolution 1.3\,cm data.

\section{Results} \label{sec:results}

\subsection{Properties of the Radio Continuum Emission} \label{sec:prop-radio-cont}

Here we describe the measured properties of the radio continuum
emission in II~Zw~40 and quantify the young, still obscured, massive
cluster population in this galaxy.  We start by using the radio
continuum emission spectrum of the central region of II~Zw~40 to
determine whether the radio continuum emission there is predominantly
thermal free-free emission produced by young massive clusters or
non-thermal synchrotron emission produced by supernovae
(\S\ref{sec:radio-spectrum-ii}).  We then identify discrete radio
continuum sources and measure their flux densities
(\S~\ref{sec:radio-source-prop}). Finally, we estimate the physical
properties of the sources (\S~\ref{sec:phys-prop-radio}).

\subsubsection{Radio Continuum Spectrum} \label{sec:radio-spectrum-ii}

The radio continuum spectrum reveals the dominant emission mechanisms
in a region. In general, the radio continuum spectrum consists of two
components: a thermal component generated by free-free emission and
associated with young massive clusters and a non-thermal component
generated by synchrotron emission and associated with supernovae. To
first order, the intrinsic spectra of the optically thin free-free
emission and synchrotron emission can both be represented by power
laws with spectral indices\footnote{We define the spectral index
  ($\alpha$) as $S \propto \nu^{\alpha}$.}  of $-0.1$ and $\approx
-0.7$, respectively. The spectral index of the free-free emission is
well-defined, but the spectral index of the synchrotron emission has a
large variation because it depends on the cosmic ray electron spectrum
\citep{2005AN....326..414B}.

We measured the radio continuum spectrum of the central region of
II~Zw~40 to determine the main emission mechanisms at work
(Table~\ref{tab:radio_sed}). The flux densities at each frequency were
determined using an aperture 6\arcsec\ in diameter; this aperture
includes the entire central emission region at all three
frequencies. We estimate that the flux density errors are 20\%. This
error takes into account the typical uncertainty in absolute flux
density calibration for the VLA \citep[$\lesssim
3\%$;][]{2012arXiv1211.1300P} and the uncertainty in the photometry
derived from a comparison of different photometry methods (Gaussian
fitting, SURPHOT (see \S~\ref{sec:radio-source-prop}),
CASA\footnote{\url{http://casa.nrao.edu}} image viewer regions).

The matched {\em u-v} coverage images are not sensitive to emission on
scales larger than 18.4\arcsec\ due to the inner 11.2~k$\lambda$
cut-off in {\it u-v} space and thus may underestimate the total amount
of flux in the region.  We use the radio continuum spectrum model from
\citet{1991A&A...246..323K} to estimate the amount of missing flux in
our images. This model is based on single-dish data and interferometer
measurements that are sensitive to much larger angular scales than our
data. Comparing our data with this model shows that our matched {\em
  u-v} coverage images do not include approximately 50-60\% of total
radio continuum flux (Table~\ref{tab:radio_sed}). Therefore, the radio
continuum spectrum that we derive is specific to the compact central
region of II~Zw~40 rather than the entire galaxy and represents only
the emission with spatial frequencies between 11.2 and
1000~k$\lambda$.

Since there are flux density measurements at only three frequencies,
the most appropriate model for the data is the following
\citep{1988AJ.....96...81D,1997A&A...322...19N}:
\begin{equation}
\frac{S}{S_0} = p_{th} \left(\frac{\nu}{\nu_0}\right)^{-0.1} 
+ (1 - p_{th} )  \left(\frac{\nu}{\nu_0}\right)^{\alpha_{nt}} 
\end{equation}
where $S$ is the flux density at frequency $\nu$, $S_0$ is the flux
density at frequency $\nu_0$, $p_{th}$ is the fraction of thermal
emission, and $\alpha_{nt}$ is the spectral index of the non-thermal
emission.

\begin{figure}
\centering
\includegraphics[width=\columnwidth]{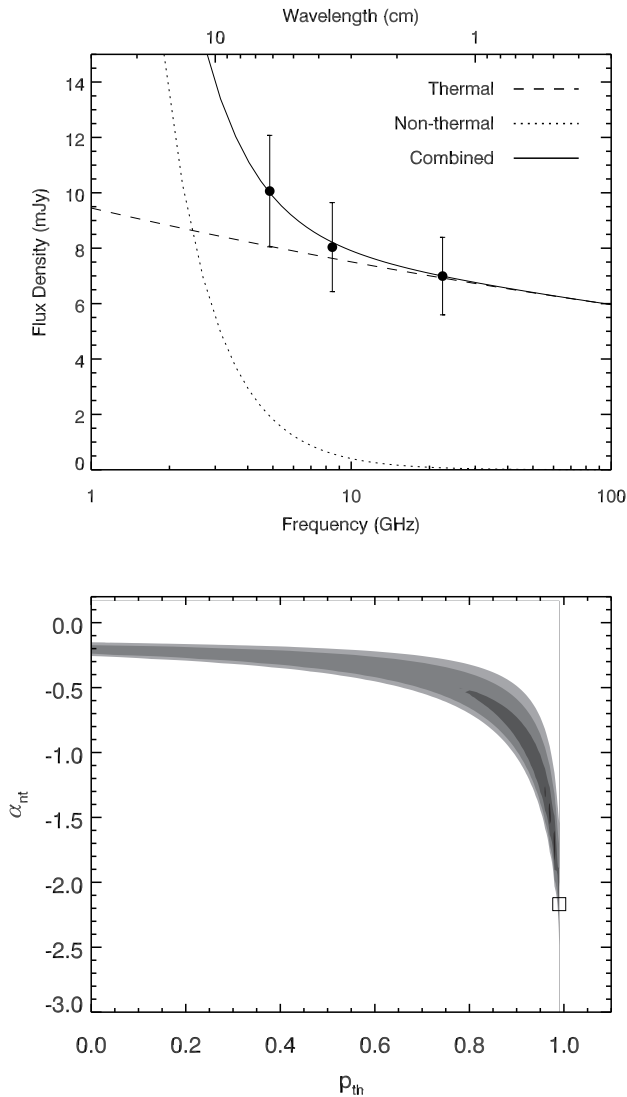}
\caption{{\em Top:} Radio continuum flux densities in the region
  corresponding to the $3\sigma$ contour in the 1.3\,cm matched beam
  image as a function of frequency. The best combined power law fit is
  shown as the solid line. The thermal contribution to the fit is
  shown as a dashed line and the non-thermal contribution to the fit
  is shown as a dotted line. The error bars on the radio continuum
  flux densities represent the 20\% uncertainty on the photometry
  rather than the formal error on the flux density. {\em Bottom:} The
  $\chi^2$ value as a function of the thermal fraction ($p_{th}$) and
  ($\alpha_{nt}$). The regions from darkest to lightest show 5, 10,
  20, and 30 times the minimum $\chi^2$ value. The open square shows
  the location of the best fit value.}
\label{fig:freq_vs_flux}
\end{figure}

The radio continuum spectrum for the central region of II~Zw~40
appears to be dominated by thermal emission with a smaller non-thermal
component (Figure~\ref{fig:freq_vs_flux}). The 1.3\,cm and 3.5\,cm
points can be fit by a single power law with a spectral index of
-0.1. Adding the 6.2\,cm data point requires including a non-thermal
component.  This non-thermal component contributes approximately 20\%
of the total flux density at 6.2\,cm.  Using the calculated $\chi^2$
values for a range of $p_{th}$ and $\alpha_{nt}$ values, the best fit
values for $p_{th}$ and $\alpha_{nt}$ are 0.99 and -2.17 with a
reduced chi-squared value of 0.0065 (bottom panel of
Figure~\ref{fig:freq_vs_flux}). The best reduced chi-squared value is
significantly lower than 1.0, suggesting that the errors on our data
may be overly conservative.  However, we retain the larger 20\% errors
because they better reflect the systematic uncertainities in our data,
not just the much smaller measurement errors.  The value for $p_{th}$
is well determined and lies between 0.75 and 1.0. The value of
$\alpha_{nt}$ is much less well determined and ranges from -0.5 to
-2.17. The latter value is unrealistically steep for $\alpha_{nt}$, so
we interpret this constraint as $\alpha_{nt} \leq -0.5$. These results
are consistent with previous studies of II~Zw~40
\citep{1986ApJ...302..640S,1991A&A...246..323K,1993ApJ...410..626D,2002AJ....124.2516B},
although the spectrum measured here is specific to the spatial scales
included in our images and is not be representative of the total radio
continuum spectrum of II~Zw~40 because of the missing large-scale
flux.

The weak synchrotron emission in the central region of II~Zw~40
implies its current supernova rate is low. In our Galaxy, the
synchrotron luminosity is related to the supernova rate by
\begin{equation}
\left( \frac{L_N}{10^{22} {\rm W \, Hz^{-1}}} \right) \sim 13 \left(
\frac{\nu}{{\rm GHz}} \right)^{\alpha_{nt}} \left( \frac{\nu_{sn}}{yr^{-1}}
\right)
\end{equation}
where $L_N$ is the supernova luminosity at frequency $\nu$ and
$\nu_{sn}$ is the supernova rate
\citep{1992ARA&A..30..575C}.\footnote{\citet{1992ARA&A..30..575C} use
  the opposite spectral index convention as in this paper. We have
  rewritten this equation to reflect the convention used here.} Using
$\alpha_{nt}=-0.5$, we estimate a current supernova rate in the center
of II~Zw~40 of $4\times10^{-4} \ {\rm yr^{-1}}$, which is accurate
within a factor of a few. 

Supernova rates from Starburst 99 models are similar to this value for
continuous bursts with ages less than 3.5~Myr and for instanteous
bursts with ages either less than 3.5~Myr or greater than 40.0~Myr
\citep{1999ApJS..123....3L}. These limits reflect the upper and lower
stellar mass limits for supernovae in Starburst~99, which are
100$\Msun$ and 8$\Msun$. At ages less than 3.5~Myr, the 100$\Msun$
stars have not evolved enough to go supernova and at ages greater than
40~Myr, all the stars with masses greater than 8$\Msun$ have already
exploded. For the continuous burst, only the upper mass limit matters
because new clusters are continually being produced.

The age of the burst seen in the radio continuum is most likely less
than 3.5~Myr. Based on optical and near-infrared photometry and
spectra, \citet{1996ApJ...466..150V} estimate that the starburst is
less than 4~Myr old.  II~Zw~40 also shows other signs of a recent
burst of star formation: optically visible super star clusters (SSCs)
and several ultra compact \hii\ regions, which may harbor young, still
obscured, SSCs
\citep[][]{2002AJ....124.2516B,2008A&A...486..393V}. However, our
radio continuum age estimate relies on the relationship between the
observed synchrotron emission and the supernova rate being roughly the
same in II~Zw~40 as in the Milky Way.  The escape of cosmic ray
electrons from II~Zw~40 or cosmic ray energy losses in its dense
starburst environment \citep{2010ApJ...717....1L} could reduce the
amount of observed synchrotron emission per supernova.

In contrast to previous results by \citet{2002AJ....124.2516B}, we do
not find any evidence for discrete sources with purely positive
spectral indices (and thus optically thick thermal emission) at the
size scales probed by our matched beam images.  This contrast is
consistent with a difference in the adopted 1.3\,cm flux scales; Beck
et al.\ adopt a 1.3\,cm flux scale a factor of 1.34 greater than the
1.3\,cm flux scale adopted here.  Using a flux density scale that is
too large would increase the derived flux at 1.3\,cm and give the
appearance of a positive spectral index at 1.3\,cm.  Our flux density
scale calibration is more robust than the flux density scale
calibration in Beck et al.\ because we use a model for the flux
calibrator source structure rather than assuming it is a point source.
These models were not available when the data used in Beck et al.\
were calibrated. For high angular resolution data, this step is
critical because the flux calibrator sources are unlikely to be point
sources. We confirmed that the difference between the radio continuum
spectrum in Figure~\ref{fig:freq_vs_flux} and the radio continuum
spectrum in \citet{2002AJ....124.2516B} is not due to differences in
the {\em u-v} range imaged; we created a set of images using our data
with the same {\em u-v} cutoff as in Beck et al.\
($\gtrsim$20k$\lambda$) and did not see evidence for any sources with
positive spectral indices.

\begin{figure}
\centering
\includegraphics[width=\columnwidth]{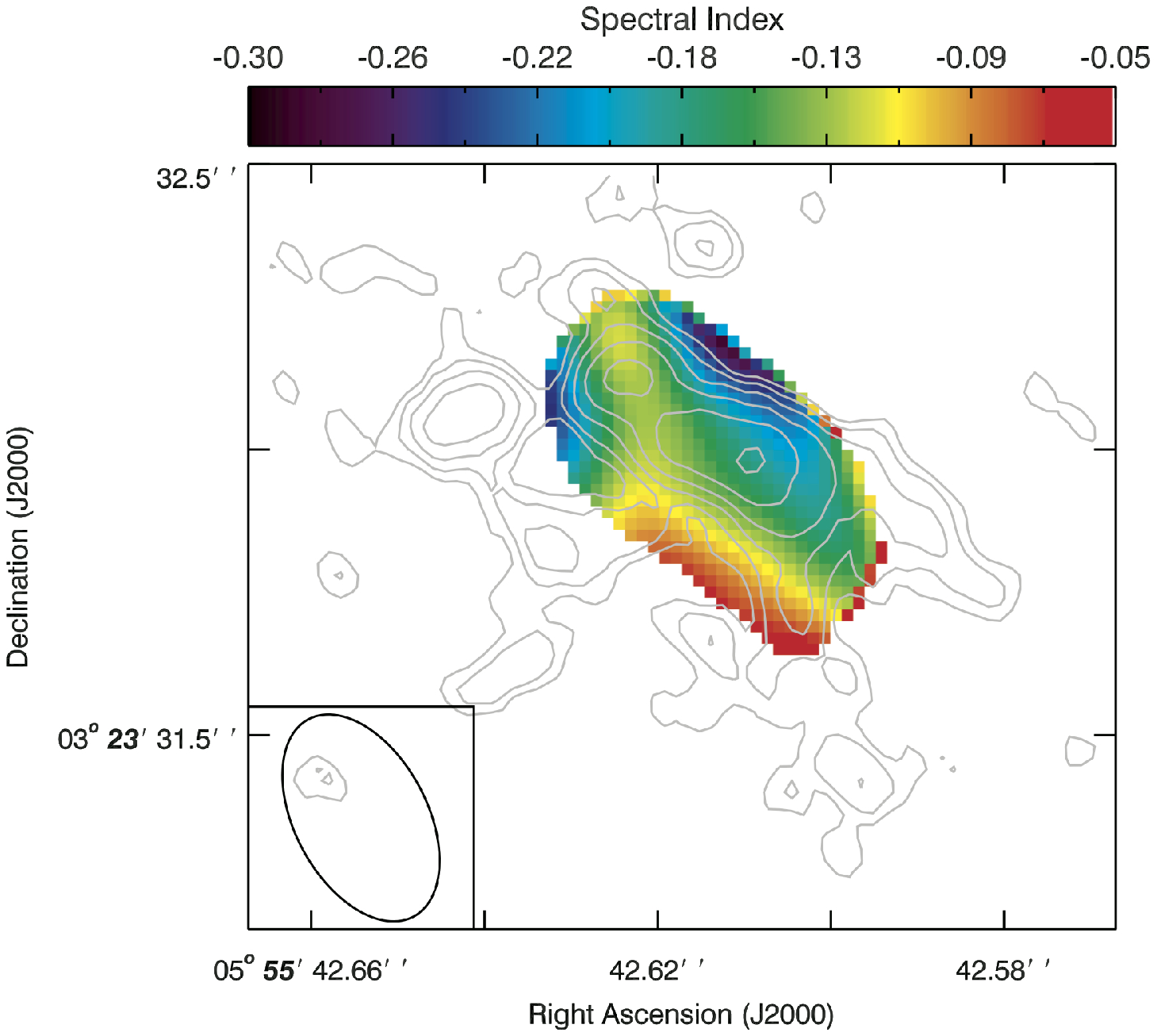}
\caption{Spectral indices between 6.2\,cm and 1.3\,cm matched {\em
    u-v} coverage images are consistent with optically thin, free-free
  emission with a small contribution from synchrotron. There are no
  purely synchrotron sources. Pixels with spectral index errors
  greater than 0.05 were blanked.  The 1.3\,cm, robust=0 contours are
  is overlaid. }
\label{fig:IIzw40_ck_spidx_kband_r0_contours}
\end{figure}

We also created a spectral index map to explore the properties and
morphology of the radio continuum emission on smaller scales
(Figure~\ref{fig:IIzw40_ck_spidx_kband_r0_contours}). Spectral index
maps are sensitive to systematic errors. Both of the input images {\em
  must} sample the same spatial scales and have the same
resolution. Therefore, we cannot use the high resolution, robust=0,
1.3\,cm image to create the spectral index maps because the maps at
other frequencies would not match its resolution or include the same
range of spatial scales.  Instead, we created a spectral index map
using the matched {\em u-v} coverage images (see \S~\ref{sec:vla-data}
for details). We blanked any pixels where the formal error on the
spectral index was greater than 0.1.  While the resolution of these
images is two to three times larger than the robust=0, 1.3\,cm images,
it provides a general picture of the spectral indices for different
parts of the central star-forming region of II~Zw~40.

As one might expect from the radio continuum spectrum, the spectral
index map shows that the central region of II~Zw~40 is dominated by
optically thin free-free emission with a contribution from synchrotron
emission (Figure~\ref{fig:IIzw40_ck_spidx_kband_r0_contours}).  The
indices in this map vary between -0.26 and -0.08. The contribution
from synchrotron emission appears to increase towards the northwestern
and the northeastern edges of the galaxy, but there do not appear to
be any discrete synchrotron dominated ($\alpha \lesssim -0.7$) sources
similar to the candidate AGN in the dwarf starburst galaxy He 2-10
\citep{2011Natur.470...66R,2012ApJ...750L..24R} or supernova remnants.

\subsubsection{Identifying Radio Continuum
  Sources} \label{sec:radio-source-prop}

In addition to constraining the dominant continuum emission mechanisms
in II~Zw~40 (\S\ref{sec:radio-spectrum-ii}), these radio continuum
data are well-suited to measuring the properties of the young massive
clusters in this galaxy. The goal of this subsection is to use the
robust=0, 1.3\,cm radio continuum image with a linear resolution of
$\sim$7~pc to identify discrete radio continuum sources (which may
consist of one or more clusters) and measure their flux densities.

The discrete sources in the robust=0, 1.3\,cm image were identified by
examining contour plots overlaid on greyscale images of the
emission. A source was identified where the presence of a $6\sigma$ or
greater contour defined a bound region. The sources are shown in
Figure~\ref{fig:kband_hst_4panel}. We do not have data with sufficient
resolution at 3.5 and 6.2\,cm to determine the radio continuum
spectral index for each source and conclusively identify the emission
mechanism. However, we expect that most of the discrete sources
identified at 1.3\,cm are thermal because the radio continuum spectrum
is dominated by free-free emission at that wavelength
(\S~\ref{sec:radio-spectrum-ii}).

To quantify this, we compare the noise in our robust=0, 1.3\,cm with a
fiducial \hii\ region (W49) and a fiducial supernova remnant (Cas
A). We estimate a 22.46~GHz flux density for a W49A-like region of
0.1~mJy by extrapolating from the measurements of
\citet{1967ApJ...150..807M} assuming strictly thermal emission and
accounting for the distance to II~Zw~40; this flux is well above our
detection limit. Cas A -- the brightest Milky Way supernova remnant --
would have a flux density of 19~$\mu$Jy at the distance of II~Zw~40,
based on the flux density models of \citet{1977A&A....61...99B}. This
flux density is below our noise limit in the 1.3\,cm, robust=0 image
(24.01~\uJybeam).  However, the upper end of the supernova remnant
luminosity function may be better sampled in galaxies with high star
formation rates \citep{2009ApJ...703..370C}. For example, the dwarf
starburst galaxy NGC 4449 has a supernova remnant with a luminosity
five times that of Cas~A \citep{2009AJ....137.3869C}. A supernova
remnant with this luminosity would be detected at only the 4$\sigma$
level in our data. Given that only one of the 43 supernova remnants in
the \citet{2009AJ....137.3869C} sample was brighter than Cas A, it is
unlikely that any of our sources are supernova remnants, even if they
are as bright as the brightest supernova remnant in NGC 4449.

The complex morphology of the thermal emission in II~Zw~40 makes
interpretation of the sources difficult. To better characterize the
radio continuum sources, we have chosen to divide our source list into
two categories: compact or diffuse. We define compact sources as
sources whose contours follow the same approximate shape down to the
sensitivity limit of the data, while the diffuse sources are the
sources whose contours significantly change shape between the highest
and lowest contour. While the compact sources (D, E, G, H, I, K) are
most likely clusters, the nature of the diffuse sources (A, B, C, F,
J) is more uncertain.  These sources could be produced by a number of
emission mechanisms including a star cluster that is superimposed on
diffuse thermal emission, the limb of an ionized gas bubble, diffuse
synchrotron emission from a supernova, or the ionized outer edge of a
molecular cloud.

\citet{2002AJ....124.2516B} identify four compact sources in
II~Zw~40. Their source at $5^{\rm{h}}55^{\rm{m}}42\fs615$,
$03\arcdeg23\arcmin32\farcs02$ (J2000) is source G in this paper. The
other three sources identified in that study, which were marginal
3.5$\sigma$ detections, do not correspond to any of the sources
identified in this paper.  We suggest that the three low signal to
noise detections identified by \citet{2002AJ....124.2516B} are
spurious given the robust=0, 1.3\,cm image used in our analysis has
comparable resolution to the uniformly weighted Beck et al.\ image and
1/7 the noise.

We measured 1.3\,cm radio continuum flux densities
(Table~\ref{tab:source_fluxes}) for each source using the program
SURPHOT \citep{2008AJ....135.2222R}, which determines the source flux
densities in irregularly shaped regions by summing the flux densities
in a user-defined region and subtracting an appropriate background. We
used the lowest contour for each source to define the regions. The
background for each source was determined using an annulus with the
same shape as the source region but a larger size. The size and width
of the annulus was chosen for each source to avoid the main body of
radio continuum emission in the region and thus accurately reflect the
background.

We can compare the global radio continuum emission in the center of
II~Zw~40 with the emission from the discrete, cluster-sized sources to
determine what fraction of the radio continuum emission originates
from young massive clusters and what fraction is more diffuse. We find
only 44\% of the total flux in the central region originates from
individual, cluster-sized sources; a large fraction of the radio
continuum emission in the center of II~Zw~40 is diffuse
(Table~\ref{tab:source_fluxes}).  The diffuse emission in the center
of II Zw 40 could be due to a porous ISM allowing ionizing photons to
escape the regions in which they were produced.  Similar fractions of
diffuse emission are seen in the low-metallicity, dwarf, starburst
galaxy SBS0335-052 \citep{2009AJ....137.3788J}. Haro 3 also exhibits
evidence for a porous and clumpy interstellar medium
\citep{2004AJ....128..610J}.

\subsubsection{Physical Properties of the Radio Continuum Sources} \label{sec:phys-prop-radio}

Given the likelihood that the discrete sources identified here are
predominantly thermal at 1.3\,cm, we derive the properties of the
embedded ionizing clusters using the 1.3\,cm continuum flux
densities. These properties include an estimate of the number of
ionizing photons produced by the young massive stars, the number of
equivalent O-stars, the cluster masses for the individual sources, and
the star formation rates for the combined sources
(Table~\ref{tab:source_properties}).

The ionizing photon flux can be estimated using
\begin{align} \label{eq:nion}
N_{ion} \ge 7.56 \times 10^{49} \left( \frac{d}{{\rm Mpc}}\right)^2 
\left( \frac{1}{1 + (n(He^+)/n(H^+))} \right) \nonumber \\ 
\times \left( \frac{\nu}{{\rm
    GHz}} \right) ^{0.1} \left( \frac{T}{10^4 \ {\rm K}}
\right)^{0.31} \left(\frac{f_\nu}{{\rm mJy}} \right)
\end{align}

where $N_{ion}$ is the number of ionizing photons, $n(He^+)/n(H^+)$ is
the ratio of ionized helium to ionized hydrogen, $\nu$ is the
frequency of the emission, $T$ is the temperature of the region, and
$f_\nu$ is the flux density of the thermal radio continuum emission
\citep{2004ApJ...606..853H}. This expression is valid for temperatures
between 10,000~K and 20,000~K. We assume a temperature for II~Zw~40 of
13,000~K, which is the average temperature measured by
\citet{1993MNRAS.262...27W}. We use a value of 0.062 for the
$n(He^+)/n(H^+)$ ratio; this is the average of the $n(He^+)/n(H^+)$
ratios found by \citet{1993MNRAS.262...27W}. The resulting number of
ionizing photons inferred for each region is given in
Table~\ref{tab:source_properties}.  The number of equivalent O-stars
in Table~\ref{tab:source_properties} was calculated from this number
by assuming that one O7.5V star produces $10^{49}$ ionizing photons
per second
\citep{1990ApJS...73....1L,1994ApJ...421..140V,1996ApJ...460..914V}. The
cluster masses for each source were estimated using the number of
ionizing photons for Starburst 99 models of a instantaneous burst of
star formation with a metallicity of Z=0.004 and an age less than
3.5~Myr.

The number of ionizing photons calculated using Equation~\ref{eq:nion}
is a lower limit. Radio continuum emission is not affected by
extinction along the line of sight from the emitting medium. However,
there are several other effects that could reduce the number of
ionizing photons estimated from the radio continuum.  First, ionizing
photons could heat the dust rather than ionizing hydrogen, although we
expect this effect to be less important for lower metallicity galaxies
because of their lower dust-to-gas ratios.  For the LMC, 30\% of the
photons heat the dust rather than ionizing hydrogen
\citep{2001AJ....122.1788I}. This fraction represents an upper limit
on the fraction of photons that go into heating dust since the LMC has
a metallicity of 1/3 solar \citep[$12 + \log({\rm O/H}) =
8.3$;][]{1984IAUS..108..353D}, which is slightly higher than the
metallicity of II~Zw~40 (1/5 solar). In addition, ionizing photons can
leak out of the region if the interstellar medium is clumpy and
porous. Approximately 25\% of the photons in a region are lost to this
effect \citep{1997ApJ...475...65H,2012MNRAS.423.2933R}. Finally, if
the \hii\ region is optically thick
\citep{1999ApJ...527..154K,2003ApJ...597..923J}, then the number of
ionizing photons inferred only represents the free-free emission from
the surface of the region, rather than the entire three dimensional
volume. There is no evidence for this effect on the size scales probed
by these observations of II~Zw~40, which would manifest as regions
with positive spectral indices, indicating self-absorption. Given the
possibilities of absorption and leakage of ionizing photons, the
overall intrinsic ionizing photon fluxes estimated here could be
underestimated by as much as a factor of 2. This value is consistent
with the observed ratio of the diffuse to discrete emission determined
in \S~\ref{sec:radio-source-prop}.

The cluster properties estimated above reveal that six out of the
eleven sources identified here require fewer than 100 O-stars to
produce the measured ionizing photon fluxes. These ionizing photon
fluxes correspond to cluster masses of less than $\sim2.5 \times 10^4
\ \Msun$ \citep{1999ApJS..123....3L}. The relatively low number of
O-stars in the sources suggests that the star formation in the center
of II~Zw~40 is not concentrated in a single massive cluster, but is
distributed across the region in several less massive clusters. In
addition, there are many extended, faint features, which suggest a
porous and clumpy ISM rather than a density- or radiation-bounded
region.

Compared to the closest example of a starburst region, 30 Doradus in
the Large Magellanic Cloud, the center of II~Zw~40 is the same size as
30 Doradus' central cluster NGC 2070 ($\sim$40~pc across
\citep{1991IAUS..148..145W}), but has an ionizing photon flux six
times higher \citep{1978MNRAS.185..263M}. The individual radio
continuum sources in II~Zw~40 are similar in size to the core of R136
\citep[2.5~pc;][]{1994AJ....107.1054M}, but three of these sources
have ionizing photon fluxes greater than or equal to that of R136 (D,
G, and H). All three of these sources are also classified as compact
sources, which strengthens our identification of the compact sources
as probable young massive clusters. Overall, the center of II~Zw~40
appears to be a more extreme starburst than 30~Doradus.

The star formation rates for the combined sources in II~Zw~40 were
estimated from the ionizing photon flux using
\begin{equation} \label{eq:sfr}
\left( \frac{SFR}{M_\sun \ {\rm yr^{-1}}}  \right) \sim 7.269 \times
10^{-54} N_{ion} 
\end{equation}
\citep{2011ApJ...737...67M}.\footnote{Star formation rates derived
  from free-free emission are not affected by dust absorbing the
  photons emitted by the ionized gas as an \ha-based star formation
  rate would be, and thus do not need a 24\micron-correction term
  \citep{2011ApJ...737...67M}. However, the star formation rate may
  still be an underestimate because it does not account for dust
  absorption of the ionizing photons from the young massive
  stars \citep{2012AJ....144....3L}. } The star formation rate for the
entire central region of II Zw 40 is 0.34~$\Msun \, yr^{-1}$; see
Table~\ref{tab:source_properties} for the star formation rates in the
other regions.

Equation (\ref{eq:sfr}) assumes a 100~Myr constant star formation rate
and solar metallicity. However, simulations by
\citet{2007ApJ...666..870C} show that star formation rates determined
using ionizing photon fluxes only vary by 20\% for younger burst ages
and metallicities between 1 and 1/5 solar. An estimate of the effect
of a shorter burst age can be made by dividing the total cluster mass
($7.4 \times 10^5$ \Msun) by the upper limit on the age of the burst
\citep[3.5~Myr;
\S~\ref{sec:radio-spectrum-ii};][]{1996ApJ...466..150V}. This
calculation gives a minimum star formation rate of 0.2~$\Msun \,
yr^{-1}$, which is consistent with the star formation rate estimates
derived above. Finally, we note that while the star formation rates
calculated above are internally self-consistent, different star
formation rate calibrations give star formation rates that vary by a
factor of 2 \citep{2012AJ....144....3L}.

The star formation rates show that star formation in II~Zw~40 is
centrally concentrated with a high surface density.  The star
formation rate for the 20~pc by 42~pc central region of II~Zw~40 is a
quarter of the total star formation rate of II~Zw~40 itself ($\sim1.4
\ {\rm M_\odot \, yr^{-1}}$; estimated using a 1.2\,cm single dish
flux density from \citealt{1984A&A...141..241K} and
Equations~(\ref{eq:nion}) and (\ref{eq:sfr})) and an eighth of the
total star formation rate of the Milky Way ($\sim 2 \ {\rm M_\odot \,
  yr^{-1}}$; \citealt{2011AJ....142..197C}). The star formation rate
surface density of the central region of II~Zw~40 is $520 \ \Msun \,
yr^{-1} \, kpc^{-2}$.  This value is twenty times higher than the star
formation rate surface density of the extremely low-metallicity (1/50
solar) dwarf starburst galaxy SBS0335-052 \citep{2009AJ....137.3788J}
and corresponds to the star formation rate surface density values seen
in the most strongly starbursting galaxies in
\citet{2008AJ....136.2846B}. The star formation rate surface density
of II~Zw~40 is greater than the empirical upper limit for star
formation rate surface densities ($45~\ \Msun \, yr^{-1} \, kpc^{-2}$)
derived by \citet{1997AJ....114...54M}.  However, it is less than the
upper Eddington limit on the star formation rate surface density set
by the effect of radiation pressure on dust \citep[$1000~\ \Msun \,
yr^{-1} \, kpc^{-2}$;][]{2005ApJ...630..167T}. Therefore, we do not
expect a radiation-driven outflow of material from II Zw 40.

\subsection{Optical Cluster Properties}\label{sec:prop-optic-clust}

A complete picture of the recent star formation in the center of
II~Zw~40 includes both the properties of the obscured clusters derived
from radio continuum in \S~\ref{sec:prop-radio-cont} and the
properties of the two unobscured optical clusters seen in the {\em
  HST} images (Figures~\ref{fig:overview} and
\ref{fig:kband_hst_4panel}). We refer to the two dominant optical
clusters as SSC-North and SSC-South.  SSC-North is surrounded by
bright ionized gas detected at both optical and radio wavelengths,
while SSC-South is not associated with strong nebular emission. This
section presents estimates of the number of ionizing photons, masses,
and ages of the optical clusters from measurements of their optical
flux densities.

Optical flux densities of SSC-North and SSC-South were obtained using
SURPHOT \citep{2008AJ....135.2222R}.  We used a circular aperture of
radius 7 pixels (0\farcs175) which includes the first Airy ring of the
PSF.  For SSC-South, the background was estimated in an annulus with
inner and outer radii of 12 and 18 pixels (0\farcs3 and 0\farcs45).
For SSC-North, the background was estimated in a more extended annulus
with inner and outer radii of 18 and 28 pixels (0\farcs45 and
0\farcs70) to avoid the extended ionized gas emission not directly
associated with the cluster.  Aperture corrections were not applied
since they are negligible for such a large aperture as determined
using the Tiny Tim web
interface.\footnote{\url{http://www.stsci.edu/software/tinytim/tinytim.html}}
The derived flux densities for the clusters are given in
Table~\ref{tab:optical_fluxes_hst}. The ionizing photon fluxes and
number of equivalent O stars was calculated from the foreground
Galactic extinction corrected \ha\ flux using Equations
(\ref{eq:nion}) and (\ref{eq:f_ha_est}) and an E(\bv)=0.57
\citep{1982AJ.....87.1165B}.\footnote{We use the Burstein and Heiles
  E(\bv) values because the E(\bv) values from
  \citet{1998ApJ...500..525S} and its re-calibration by
  \citet{2011ApJ...737..103S} are larger than the average total
  extinction we find in the near-infrared for II Zw 40 by comparing
  our radio continuum data and near-infrared data from the literature
  (see \S~\ref{sec:galaxy-wide-extinct}).}

Physical properties of the clusters were estimated by comparing the
observed photometry to Starburst99 (Version 5.1,
\citealp{1999ApJS..123....3L}) and GALEV \citep{2009MNRAS.396..462K}
evolutionary synthesis models.  Details of the models used here can be
found in \citet{2010ApJ...708...26R}.  Briefly, both sets of models
are for an instantaneous burst\footnote{For SSC-North, the choice of
  instantaneous or continuous burst of star formation does not affect
  the derived age. For SSC-South, a continuous burst of star formation
  would require an unrealistically old age for the cluster ($\gtrsim 1
  {\rm Gyr}$), given that the current burst of star formation in II Zw
  40 is less than 4 Myr old \citep{1996ApJ...466..150V}.} with a
Kroupa IMF and use the Geneva evolutionary tracks with a metallicity
of $Z=0.004$ to match the measured metallicity of II~Zw~40
\citep[$≈ƒ12 + \log({\rm O/H}) = 8.09$;][]{2000ApJ...531..776G}.
Stellar and nebular continuum are included in both sets of models,
however, the GALEV models also include emission lines (e.g.\
\citealp{2003A&A...401.1063A}).

Ages of the clusters were estimated using the equivalent width (EW) of
H$\alpha$.  The EWs were calculated as $(f_{F658N} - f_{cont} ) \Delta
\lambda_{F658N} / f_{cont}$ where $f_{F658N}$ is the flux density
through the F658N filter, $\Delta \lambda_{F658N}$ is the width of the
F658N filter (78\AA, {\it HST} ACS Cycle 18 Instrument Handbook, Table
5.1), and $f_{cont}$ is the continuum (stellar + nebular) flux
density.  The archival {\it HST} ACS observations did not include a
corresponding continuum filter for the F658N filter. Therefore, the
continuum was estimated by interpolating between the flux densities in
the adjacent F555W and F814W filters. The H$\alpha$ EWs of SSC-North and
SSC-South are $\sim 460$\AA\ and $\sim 50$\AA, respectively.  The
corresponding ages inferred from the Starburst99 models are $\lesssim
5.0$ Myr for SSC-North and $\sim$9.4$^{+1.4}_{-1.0}$~Myr for
SSC-South.

The presence of emission lines in the F555W and F814W filters may
affect the estimated continuum and thus the \ha-derived optical
cluster properties. While SSC-South is old enough that line
contamination is minimal, SSC-North is young enough that line
contamination could significantly affect the continuum estimate (see
Figure 11 in \citealp{2010ApJ...708...26R}). We estimate the line
contamination for the SSC-North F555W and F814W measurements using the
line contamination calculated in Table 4 of
\citet{2010ApJ...708...26R} for a cluster in the galaxy NGC 4449 with
a similar age and metallicity. From this table, we find that line
emission could contribute about 25\% of the flux in the F814W filter
and about 63\% of the flux in the F555W filter for SSC-North. After
correcting the flux for SSC-North in these filters for the estimated
line contamination, the resulting \ha\ fluxes (and number of ionizing
photons and number of O stars) only differ by 10\% from the estimates
given in our Table~\ref{tab:source_properties_hst}. The resulting \ha\
equivalent width changes from $\sim460$\AA\ to $\sim986$\AA, but age
limit derived from this quantity does not change significantly
($\lesssim 4 {\rm Myr}$ instead of $\lesssim {\rm 5 Myr}$).

\begin{figure}
\centering 
\includegraphics[width=\columnwidth]{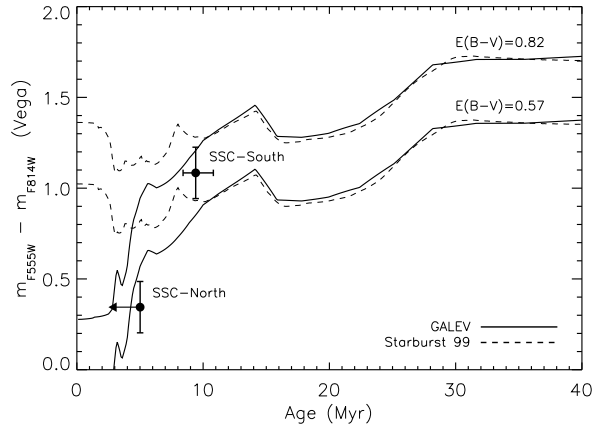}
\caption{The north and south optical clusters in II~Zw~40 are young
  with ages of $\lesssim5.0$~Myr and $9.4^{+1.4}_{-1.0}$ Myr,
  respectively. The ages were determined using the \ha\ equivalent
  width (see \S~\ref{sec:prop-optic-clust} for details). The solid
  lines show the GALEV models that include emission lines in addition
  to nebular continuum and the dashed lines show the Starburst99
  models that only include nebular continuum. The two optical clusters
  are indicated. We show models using the \citet{1982AJ.....87.1165B}
  extinction, E(B-V)=0.57, and the \citet{1998ApJ...500..525S}
  extinction, E(B-V)=0.82. } \label{fig:color_evolution}
\end{figure}

To estimate the extinction of the clusters, we compared the observed
$m_{F555W}-m_{F814W}$ color to the Starburst99 and GALEV model colors.
Figure~\ref{fig:color_evolution} shows the model colors as a function
of age, accounting for Galactic foreground reddening using the
extinction curve of \citet{1989ApJ...345..245C}.  Color evolutionary
tracks using $E(B-V)=0.82$ \citep{1998ApJ...500..525S} and
$E(B-V)=0.57$ \citep{1982AJ.....87.1165B} are shown for both the
Starburst99 and GALEV models.  Both clusters appear to have minimal
internal reddening.  However, it is possible that some of the optical
light is completely absorbed (and not impacting our measured
extinction) in a dense and clumpy ISM
\citep[e.g.][]{2008AJ....136.1415R}, especially for the younger
SSC-North.
  
Masses of the clusters were estimated by comparing the observed flux
densities in the F555W and F814W filters to the model flux densities
for a $10^5$ M$_\odot$ cluster.  Lower limits of the cluster masses
were obtained by assuming no internal extinction and a Galactic
foreground reddening of $E(B-V)=0.57$.  For SSC-North, we used the
GALEV model since emission lines are significant at such a young age
\citep[e.g.,][]{2010ApJ...708...26R} and obtained a minimum mass of $0.9
\times 10^5$ M$_\odot$.  For SSC-South, we obtained a minimum mass of
$1.6 \times 10^5$ M$_\odot$.  The GALEV and Starburst99 models give
nearly the same result for this cluster since the relative importance
of nebular emission is significantly diminished by $\sim10$
Myr. Table~\ref{tab:source_properties_hst} gives an overview of the
derived properties of the optical clusters.

The masses and ages of the optical clusters derived here agree with
other estimates of their age and mass.  Using Br$\gamma$ observations
of SSC-North, \citet{2008A&A...486..393V} derive an upper limit on its
age of 3~Myr, which agrees with the age derived above. Using our data
and the same size aperture as Vanzi et al.\ (15\arcsec\ instead of
0.35\arcsec), we derive a mass for SSC-North of $(1-5) \times
10^6$~M$_\odot$, which is also in agreement with their mass estimate
of $1.7 \times 10^6$ M$_\odot$.  For SSC-South, Vanzi et al.\ find an
age of 6 to 7 Myr and a mass of $1.3 \times 10^5$ M$_\odot$, which are
similar to our derived values.

\begin{figure}
\centering
\includegraphics[width=\columnwidth]{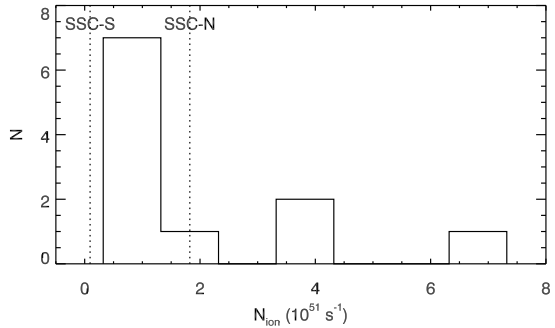}
\caption{In the center of II~Zw~40, the optical clusters have ionizing
  photon fluxes similar to the fainter radio continuum sources. The
  histogram shows the radio continuum ionizing photon fluxes
  determined in \S~\ref{sec:radio-source-prop}. The ionizing photon
  fluxes for the optical clusters from \S~\ref{sec:prop-optic-clust}
  are marked with dotted lines and labeled with the cluster name. The
  ionizing photon flux from SSC-N is a lower limit.}
\label{fig:photon_budget}
\end{figure}

A comparison of the optically visible clusters and the radio sources
shows that the optical clusters have ionizing photon fluxes (derived
from \ha) similar to the fainter radio continuum sources
(Figure~\ref{fig:photon_budget}).  The brightest 3 radio continuum
sources have 2-4 times more ionizing photons than the brightest
optical cluster.  However, the number of ionizing photons may be
underestimated by a factor of a few for the optical clusters. First,
we have chosen an aperture including only the cluster and have
excluded much of the surrounding emission; this effect may be
particularly important for SSC-N. Second, the foreground extinction
may be underestimated; we use the \citet{1982AJ.....87.1165B} value
rather than the larger \citet{1998ApJ...500..525S} or
\citet{2011ApJ...737..103S}. However, we believe our foreground
extinction estimate is appropriate because larger values lead to more
extinction in the infrared than we measure for II~Zw~40
(\S~\ref{sec:galaxy-wide-extinct}).

The brightest clusters seen in the radio may be producing more
ionizing photons than the brighest optical cluster because of
differences in either the ages of the clusters or the mass
distribution of the clusters. As clusters age, the number of ionizing
photons decreases because the most massive stars in the cluster die
first.  \citet{1999ApJS..123....3L} shows that for a cluster of a
given mass the number of ionizing photons as a function of time is
roughly constant for the first 4~Myr of a burst. Depending on the
relative ages of the clusters seen in the radio continuum the
difference in the number of ionizing photons between the radio and
optical clusters is between 1.5 times (for a radio continuum cluster
age of 3 Myr and an optical cluster age of 5 Myr) and 5.6 times (for a
radio continuum cluster age of 1 Myr and and optical cluster age of 5
Myr). The mass distribution of the clusters may also change with time,
which would lead to differences in the ionizing photon flux between
the younger, still obscured radio cluster population and the slightly
older, less obscured optical cluster population. If the star formation
in the center of II~Zw~40 is ramping up and producing more massive
clusters, the radio should see an increase in the number of ionizing
photons because it is probing a younger generation that is producing
more ionizing photons because there are more massive clusters than in
the optical. On other hand, if the star formation rate is roughly constant, the
difference between the radio and optical sources could reflect the
destruction of clusters \citep{2009ApJ...704..453F}.  Unfortunately,
we do not have enough information here to disentangle these
possibilities.

\section{Discussion} \label{sec:discussion-1}

This section uses the properties of the radio continuum emission
(\S~\ref{sec:prop-radio-cont}), and the optical clusters
(\S~\ref{sec:prop-optic-clust}) to explore star formation in the
center of II~Zw~40. Our goal is to identify whether star formation in
this low-metallicity dwarf starburst galaxy is different than star
formation in other environments, and if so, what the differences
are. First, we quantify the properties of the young massive cluster
population in II~Zw~40 using the \hii\ region luminosity function
(\S~\ref{sec:hii-regi-lumin}). Then, we discuss the internal
extinction due to dust in II~Zw~40
(\S~\ref{sec:dust-properties}).

\subsection{The \hii\ Region Luminosity Function} \label{sec:hii-regi-lumin}

In this section, we derive the \hii\ region luminosity function for
the central region of II~Zw~40 using its radio continuum source
properties (\S~\ref{sec:radio-source-prop}).  The resolution of our
data is well-matched with the size scales of clusters allowing us to
directly probe the cluster luminosity function. We compare the derived
\hii\ region luminosity function for II~Zw~40 with \hii\ region
luminosity functions in other galaxies to determine whether the
physical conditions in II~Zw~40 may impact the cluster population.

A fundamental issue with luminosity functions in dwarf galaxies is
that there will be few sources due to the low mass of these
galaxies. However, it is still important to build up statistics on
their luminosity functions because these galaxies have physical
environments that differ from more massive systems.

\begin{figure}
\centering
\includegraphics[width=\columnwidth]{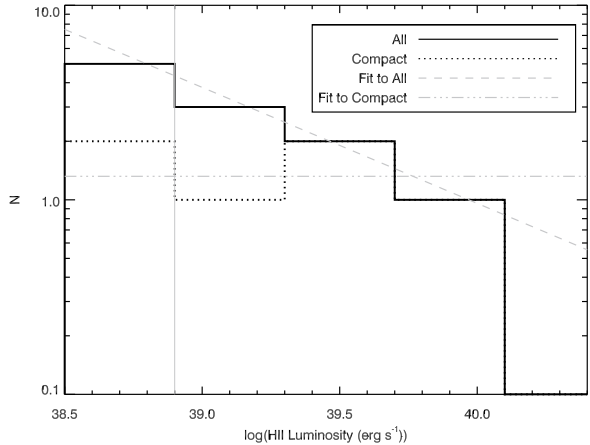} 
\caption{The slope of the \hii\ region luminosity function for the
  radio continuum sources identified in II~Zw~40 is flatter than those
  found for earlier type galaxies. The \hii\ region luminosity
  function for all sources in II~Zw~40 is shown as a solid line, while
  the \hii\ region luminosity function for just the compact sources is
  a dotted line. The fits to the luminosity functions have slopes of
  $-1.59\pm0.09$ for all sources (dash-dotted line) and $-1.00\pm0.43$
  if only compact sources are included (dash-triple dotted line).
  Both fits exclude the faintest luminosity bin, which is below the
  3$\sigma$ completeness limit (shown here as a thin, grey, vertical
  line). Earlier type galaxies tend to have steeper slopes ($\sim -2$)
  \citep{1989ApJ...337..761K}). This effect has been attributed to
  cluster sampling statistics \citep{1998AJ....115.1543O}. Later type
  galaxies tend to have brighter clusters, which tends to produce a
  flatter luminosity function. Note that fits to luminosity functions
  in $\log(N)-\log(L)$ result in an apparent slope of $\alpha+1$
  because the luminosity function is defined in linear intervals, but
  the plot is in logarithmic
  intervals.} \label{fig:cluster_luminosity_function}
\end{figure}

\hii\ region luminosity functions are typically modeled as power laws:
\begin{equation}
N(L) = A \, L^{\alpha} \, {\rm d}L
\end{equation}
where $N(L)$ is the number of clusters in the interval $L$ to $L+{\rm
  d} L$, $\alpha$ is the slope,\footnote{Unfortunately, the variable
  $\alpha$ is commonly used for both the slope of the \hii\ region
  luminosity function and the slope of the radio continuum
  spectrum. We retain the common usage of these variables to avoid
  confusion with other papers in the literature and rely on context to
  distinguish them.}  and $A$ is the normalization constant
\citep{1989ApJ...337..761K}. The fitted HII region luminosity function
slopes are $-1.59\pm0.09$ for all the sources and $-1.00\pm0.43$ for
only the compact sources
(Figure~\ref{fig:cluster_luminosity_function}). These fits exclude
sources below the completeness limit.

\begin{figure}
\centering
\includegraphics[width=\columnwidth]{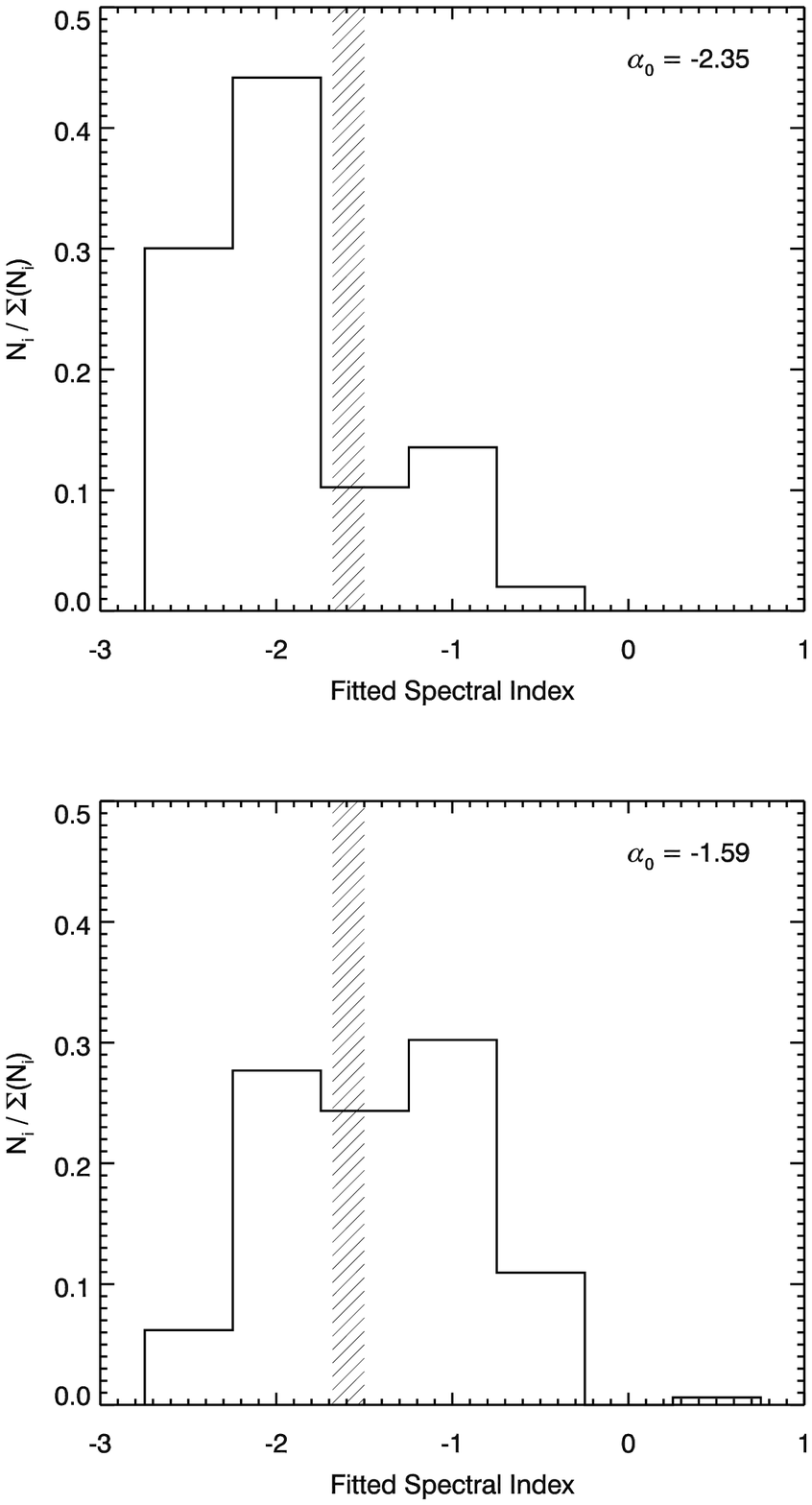}
\caption{Even with the small sample here we can determine the slope of
  the \hii\ region luminosity function. This plot shows the
  distribution of the slopes fit to simulated \hii\ region luminosity
  functions with the same number of sources as we observe. The \hii\
  region luminosity functions were drawn from two power law
  distributions: one with a slope of -2.35 and the other with a slope
  of -1.59. The fitted slope values from the Monte Carlo simulation
  tends to cluster around the true slope value. The slope value
  determined from our data is shown as the hashed region. We would
  measure this value only 7\% of the time if the true slope was
  -2.35.}
\label{fig:mc_alpha235_alpha159}
\end{figure}

Determining the slope is possible because it relies on the
distribution of all the sources rather than the number of counts in a
particular luminosity bin. To demonstrate this, we carried out Monte
Carlo simulations which generated 1000 \hii\ region luminosity
functions drawn from a power law distribution with the same number of
sources as above and fit the slope of each luminosity function. We
used two power law distributions: one with a slope of -2.35, which is
comparable to the ``typical'' slope for a cluster mass function
\citep{2012ARA&A..50..531K}, and the other with a slope of -1.59, the
slope derived above for the entire sample. Even with a small number of
sources, the distribution of fitted slopes for the simulated \hii\
region luminosity functions tends to cluster around the true slope of
the \hii\ region luminosity function
(Figure~\ref{fig:mc_alpha235_alpha159}). If the true slope were -2.35,
the data would only yield a fit with a slope of $-1.59\pm0.09$ 7\% of
the time.

The slope of the \hii\ region luminosity function tends to decrease
with the Hubble type of the galaxy; Sa/Sb galaxies have much steeper
slopes than irregular galaxies \citep{1989ApJ...337..761K}. The slopes
of the \hii\ region luminosity functions for two large spiral
galaxies, M101 and M31, are $-2.3\pm0.2$ and $-2.1\pm0.2$,
respectively \citep{1989ApJ...337..761K}, while the slope for the LMC
is $-1.75\pm0.15$ \citep{1989ApJ...337..761K}. For a sample of
irregular galaxies, the average slope is $-1.0\pm0.1$ for a sample
with no turnover in the luminosity function and $-1.5\pm0.1$ for a
sample with a turnover in the luminosity function
\citep{1999ApJ...519...55Y}.

We cannot directly compare the slope of our \hii\ luminosity function
to the slopes determined by \citet{1989ApJ...337..761K} and
\citet{1999ApJ...519...55Y} because we have much higher resolution
($\sim7{\rm pc}$) than these papers ($\sim 30 {\rm pc}$). \hii\
regions identified at low resolutions (30-70 pc) separate into
individual \hii\ regions at higher resolutions (5-10 pc) causing the
slope of the luminosity function to steepen
\citep{2000A&A...361..913P,2001AJ....122.3017S}. However, the relative
trend of flatter slopes for later type galaxies should be the same for
a data set with uniform resolution. Studies of \hii\ region luminosity
functions with 10 pc resolution are rare. In one example,
\citet{2011AJ....141..113G} derive \hii\ region luminosity functions
using high resolution {\em HST} data. They find a similar trend to
\citet{1989ApJ...337..761K}: the irregular galaxy NGC 4449 has a
flatter luminosity function slope ($\alpha=-1.43$) than the spiral
galaxy M51 ($\alpha = -1.79$).

The flat slope of the \hii\ region luminosity function in II~Zw~40 may
be due to relatively more clusters with high ionizing photon
fluxes. This could result from either more massive stars per cluster
(i.e., a top-heavy IMF) or a higher relative number of massive
clusters than in normal spiral galaxies
\citep{1989ApJ...337..761K}. The latter possibility is more likely
because the IMF in most galaxies appears to be roughly constant
\citep{2010ARA&A..48..339B}.  Stochastic models of \hii\ region
luminosity functions suggest more massive clusters in a galaxy produce
flatter luminosity function slopes
\citep{1998AJ....115.1543O}. However, the relationship between \hii\
region luminosity functions and cluster mass functions depends on the
sampling of the IMF to convert ionizing photon fluxes to masses. This
complexity makes any conclusions on the form of the cluster mass
function uncertain.

As discussed in the introduction, there are several mechanisms that
could lead to galaxies like II~Zw~40 having a disproportionate number
of more massive clusters, including: low large-scale shear in a
quiescent galaxy \citep{2010ApJ...724.1503W}, high molecular gas
surface densities leading to less effective radiation pressure
\citep{2010ApJ...713.1120K}, or high ISM pressures
\citep{1997ApJ...480..235E}. II~Zw~40 is in the end stages of a merger
as evidenced by its optical (Figure~\ref{fig:overview}) and neutral
hydrogen emission \citep{1998AJ....116.1186V}. Therefore, we conclude
that the most likely explanation for the possible presence of more
massive clusters in II~Zw~40 is high ISM pressures due to the
interaction rather than low large-scale shear, which is unlikely given
the merger history of II~Zw~40, or less effective radiation pressure,
which requires a high dust opacity.

\subsection{Internal Extinction} \label{sec:dust-properties}

This section uses the free-free emission in the radio with the \ha\
emission in the optical (as well as other hydrogen recombination lines
in the optical and infrared) to measure the internal extinction in
II~Zw~40.  Free-free emission is produced by the same electrons that
produce \ha\ emission in the optical; both of these emission
mechanisms fundamentally trace the same physical region. However,
free-free emission is not attenuated by dust, unlike \ha\ emission (or
any of the standard Balmer, Paschen, or even Brackett transitions).
Therefore, the internal dust extinction can be estimated by comparing
the \ha\ emission predicted by the free-free emission in the radio to
the observed \ha\ emission in the optical. Section
\ref{sec:indiv-source-extinct} estimates the extinctions for the
individual sources in II~Zw~40, while \S\ref{sec:galaxy-wide-extinct}
determines the internal extinction as a function of wavelength
integrated over the central region of II~Zw~40.

There are two key facts to remember about internal extinctions derived
using this method. First, the internal extinction is a convolution of
the properties of the dust {\em and} the geometry of the dust and the
young massive stars, not just the properties of the dust alone. In
other words, extinction estimates of this type are luminosity
weighted: regions with a greater contribution to the total light will
dominate.  A second related point is that this extinction estimate
will be a lower limit for the total extinction in the galaxy if
photons are absorbed by dust or escape the region rather than ionizing
hydrogen.

\subsubsection{Individual source extinction
  values} \label{sec:indiv-source-extinct}

The internal extinctions for the individual radio continuum sources
defined in \S~\ref{sec:radio-source-prop} were calculated in the
following way:
\begin{enumerate}
\item We measured the optical flux densities for each radio continuum
  source using the {\em HST} ACS data
  (Table~\ref{tab:optical_fluxes}). The continuum for the F658N filter
  was estimated using the same method as in
  \S\ref{sec:prop-optic-clust}. 
\item We corrected the optical \ha\
  fluxes for foreground Galactic extinction using
\begin{equation} \label{eq:foreground_extinction}
f_{H\alpha,c} = f_{H\alpha} 100^ {A_{gal,H\alpha}/5}
\end{equation}
where $f_{H\alpha,c}$ is the \ha\ flux corrected for foreground
Galactic extinction, $ f_{H\alpha}$ is the measured \ha\ flux, and
$A_{gal,H\alpha}$ is the foreground Galactic extinction.
(Table~\ref{tab:extinction}). The foreground extinction at the
position of II~Zw~40 is E(\bv) = 0.57 \citep{1982AJ.....87.1165B},
which translates to an extinction at the wavelength of \ha\ of 1.4
magnitudes, assuming an $R_V$=3.1 Milky Way extinction curve
\citep{1989ApJ...345..245C}.
\item  We calculated the expected \ha\ flux based on the radio continuum for
each source using the high resolution 1.3\,cm data and the following
equation:
\begin{align} \label{eq:f_ha_est} \left(
    \frac{f_{H\alpha,est}}{10^{-12} \ {\rm erg \, cm^{-2} \, s^{-1}}}
  \right) = \nonumber \\ 0.864 \left( \frac{1}{1 + (n(He^+)/n(H^+))} \right)
  \nonumber \\ \left(
    \frac{T}{10^4 \ {\rm K}} \right)^{-0.617} \left( \frac{\nu}{ {\rm
        GHz}} \right)^{0.1} \left( \frac{f_\nu}{{\rm mJy}} \right)
\end{align}
where $f_{H\alpha,est}$ is the predicted \ha\ flux and $f_\nu$ is the
flux density of the thermal radio continuum emission
\citep{2004ApJ...606..853H}. This expression is accurate for
temperatures between 10,000~K and 20,000~K and densities between 100
and 1000~${\rm cm}^{-3}$. We use the same helium fraction,
temperature, and frequency as in Section~\ref{sec:radio-source-prop}
for this calculation.
\item We calculated the internal extinction using the Galactic
  foreground corrected optical \ha\ flux and the expected \ha\ flux
  based on the radio continuum using:
\begin{equation} \label{eq:extinction_definition}
A_{H\alpha} = 2.5 \log \left(  \frac{f_{H\alpha,est}}{f_{H\alpha,c}} \right)
\end{equation}
where $A_{H\alpha}$ is the extinction in magnitudes.
\end{enumerate}

While many dwarf galaxies can be assumed to be transparent
\citep{1996ApJ...457..645W,2003ApJ...586..794B}, the internal
extinctions calculated for II Zw 40 show that this galaxy is more
opaque than the typical dwarf galaxy (Table~\ref{tab:extinction}). The
\ha\ extinction derived for the entire central region implies that
only 30\% of the ionized gas is visible in the optical. This high
level of internal extinction underscores how crucial long wavelength
observations are for studying these regions.

Other estimates of the internal extinction in II~Zw~40 from optical
hydrogen recombination lines agree with our results. They range
between $\Av = 0.23 - 1.23$ mag with a mean of approximately 0.83 mag
\citep{1970ApJ...162L.155S,1986ApJ...308..620W,1988ApJ...331..145J,1993MNRAS.262...27W,1996ApJ...466..150V,1998MNRAS.295...43D}. In
contrast, much higher values of the internal extinction are derived
from dust emission models: $\Av\ \sim 30$
\citep{2005A&A...434..849H}. The difference between the optical and
dust-derived extinctions suggests that dense gas clumps in II~Zw~40
absorb a significant number of photons that would otherwise ionize
hydrogen.  The extinction inferred from the infrared reflects the
photons that are absorbed by dust, but the extinction inferred from
the optical recombination lines can only be produced by photons
escaping between the dust clumps to ionize hydrogen, lowering the
total extinction seen along the line of sight.  This trend for
extinctions is commonly seen in starburst galaxies
\citep{1997ApJ...487..625G,2001PASP..113.1449C,2008AJ....136.1415R}
and in other dwarf starburst galaxies like Haro 3 and SBS0335-052
\citep{2004AJ....128..610J,2008AJ....136.1415R,2009AJ....137.3788J}
and is likely produced by a porous and clumpy interstellar medium.

\subsubsection{Internal Extinction Curve Integrated Over the Central Region} \label{sec:galaxy-wide-extinct}

The total internal extinction -- the internal extinction integrated
over the center of II~Zw~40 as a function of wavelength -- can be
measured by comparing our radio continuum measurements to hydrogen
recombination line measurements in the optical and near-infrared from
the literature and subtracting an estimate of the foreground Milky Way
extinction. We use the hydrogen recombination line measurements from
\citet{2011ApJ...734...82I} because these data offer uniform coverage
over a wide wavelength range (0.36 to 2.46 \micron).

We used apertures matching those used by \citet{2011ApJ...734...82I}
to measure corresponding radio continuum flux densities. The optical
spectra had an extraction aperture of $1\farcs1$ by $6\arcsec$ at a
position angle of 214$\degr$ and the near-infrared spectra had an
extraction aperture of $1\farcs5$ by $4\arcsec$ at a position angle of
128$\degr$ (Y. Izotov, 2011, private communication). The \ha\ fluxes
estimated from our radio continuum data for the optical and
near-infrared apertures are $4.76 \times 10^{-12}$ and $4.15 \times
10^{-12}$ erg~cm$^{-2}$~s$^{-1}$, respectively. We used the Case B,
100~cm$^{-3}$, 10,000~K line ratios from \citet{1995MNRAS.272...41S}
and Equation (\ref{eq:f_ha_est}) to determine the estimated fluxes for
the other hydrogen recombination lines from our radio continuum data.

The near-infrared and optical spectra in \citet{2011ApJ...734...82I}
were taken by two different instruments with two different apertures
and need to be put on the same flux density scale. Comparing the
spectra requires taking into account the distribution of the ionized
gas emission. If we assume that the radio continuum and Paschen
$\epsilon$ line have the same distribution, the flux for the Paschen
$\epsilon$ in the near-infrared aperture should be 0.935 times the
flux of that line in the optical aperture.\footnote{The internal
  extinction should not significantly change the distribution of
  Paschen $\epsilon$ emission compared to the radio continuum because
  we are considering a longer wavelength line that is much less
  affected by extinction.} The scaling used in
\citet{2011ApJ...734...82I}, which was based on the flux of the [SIII]
0.953~\micron\ line in the two spectra and did not account for the
spatial distribution of emission, produces a Paschen $\epsilon$ line
flux in the near-infrared aperture that is 0.842 times that of the
Paschen $\epsilon$ line flux in the optical aperture. Multiplying the
near-infrared values given in Table~3 of \citet{2011ApJ...734...82I}
by a factor of 1.11 reproduces the expected line flux ratio between
the Paschen $\epsilon$ line fluxes in the near-infrared and optical
apertures (0.935).

\begin{figure}
  \centering
\includegraphics[width=\columnwidth]{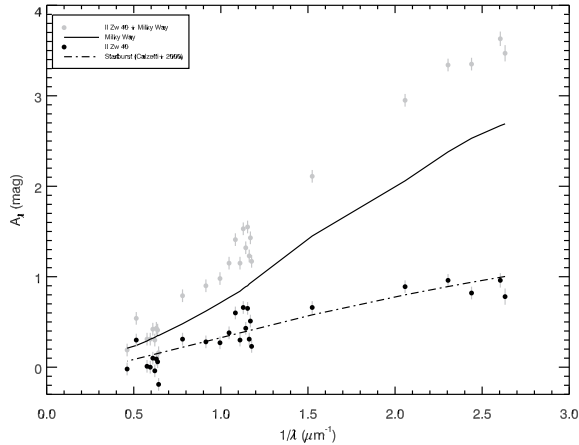}
\caption{The variation of internal extinction with inverse wavelength
  for II~Zw~40 is consistent with that of a typical starburst galaxy.
  The gray points show the total extinction (Milky Way foreground plus
  II Zw 40 internal extinction) determined by comparing the radio
  continuum emission presented in this paper with optical and
  near-infrared line fluxes from \citet{2011ApJ...734...82I}. The
  radio fluxes were measured using the same apertures as in the
  optical and near-infrared spectra. The near-infrared line fluxes
  have been scaled so that the near-infrared Paschen $\epsilon$ flux
  matches the expected near-infrared Paschen $\epsilon$ flux based on
  the distribution of the ionized gas emission and the sizes and
  orientations of the near-infrared and optical apertures. The solid
  black line shows the \citet{1989ApJ...345..245C} Galactic extinction
  curve ($R_v=3.1$, E(B-V)=0.57). The estimated internal extinction
  due solely to II Zw 40 (total extinction minus the foreground Milky
  Way extinction) is shown as black points. A fitted starburst
  extinction curve is shown as a dash-dot line
  \citep{2000ApJ...533..682C}.}
\label{fig:lambda_vs_extinction}
\end{figure}

We calculated the internal extinction for II Zw 40 as in Section
\S~\ref{sec:indiv-source-extinct} using our radio continuum data, the
re-scaled optical and near-infrared line fluxes from
\citet{2011ApJ...734...82I}, and an estimate of the foreground Milky
Way extinction (Figure~\ref{fig:lambda_vs_extinction}). The integrated
internal extinction curve for II~Zw~40 is consistent with the
metallicity-independent internal extinction curve typical of starburst
galaxies \citep{1997ApJ...487..625G}.  Therefore, the internal
extinction curve in II~Zw~40 appears to be governed by the starburst
and metallicity plays a secondary role. The internal extinction curves
for starburst galaxies are thought to be the result of the starburst
clearing the dust out of the central burst region leaving behind a
shell of clumpy dust \citep[see \S~4.2 and Figure 8
in][]{2000ApJ...533..682C}, which is consistent with our model for the
internal extinction in II~Zw~40 in \S~\ref{sec:indiv-source-extinct}.

\citet{2011ApJ...734...82I} find that II Zw 40 has similar internal
extinction in the near-infrared and the optical. Our internal
extinction curve demonstrates that, when the correct scaling factors
are applied to their data, the extinction in the near-infrared is
significantly less than the extinction in the optical.

\section{Summary and Conclusions} \label{sec:discussion}

This paper presented high resolution radio continuum observations from
the VLA and optical observations from {\em HST} of the young massive
cluster population of the prototypical dwarf starburst galaxy
II~Zw~40. These observations were used to quantify the properties of
star formation in this extreme galactic environment and compare them
to the star formation properties of other galaxies. 

We found that the radio continuum spectrum of the central portion of
II~Zw~40 is dominated by free-free emission from gas ionized by young
massive stars. The relatively weak synchrotron emission implies that
the supernova rate is low and thus that the current burst of star
formation is still relatively young ($\lesssim 3.5 \ {\rm Myr}$). The
morphology of the radio continuum distribution shows that the star
formation in the central part of II~Zw~40 is clumpy and distributed.
The total ionizing photon flux for the entire central region of
II~Zw~40 is comparable to that of NGC~2070, the central cluster
powering the 30~Doradus region in the LMC, and the star formation rate
surface density is extremely high ($520 \ \Msun \, yr^{-1} \,
kpc^{-2}$). There are 3 sources with ionizing photon fluxes comparable
to the very center of NGC~2070 (R136) in 30 Doradus, but a majority of
the sources have fewer ionizing photons. Discrete, cluster-sized
sources also only account for roughly half the emission; the remainder
is diffuse. The diffuse emission could be due to leakage of ionizing
photons from the main star-forming regions.

In the optical, there are two dominant clusters in the central
star-forming region of II~Zw~40, with the southern cluster having an
older estimated age than the northern cluster (9.4~Myr and
$\lesssim$5.0~Myr, respectively). The lower limits on the optical
cluster ionizing fluxes (which are correlated with the cluster mass)
are comparable to the ionizing fluxes of the fainter radio continuum
sources.  Given that the youngest clusters (identified in the radio)
are also the most massive implies either an evolution in the star
formation rate in the center of II~Zw~40 (i.e. star formation is still
ramping up) or an evolution of the cluster population (i.e. disruption
of the older clusters or lower ionizing photons fluxes from older
clusters).

Monte Carlo simulations verify that the \hii\ region luminosity
function is flatter than the typical cluster mass function
($\alpha=-1.59\pm0.09$ instead $\alpha \sim -2.35$), reflecting the
general trend of later type galaxies having flatter \hii\ region
luminosity functions. This trend implies that II~Zw~40 has relatively
more massive clusters than an earlier-type galaxy.  These massive
clusters may be the result of the high ISM pressures produced by the
merger that created II~Zw~40. This conclusion relies on being able to
convert ionizing photon fluxes to masses, which assumes that the IMF
is normal and well-populated.

The extinctions derived for II~Zw~40 show that, even in
low-metallicity galaxies, young massive star-forming regions are
relatively opaque. Only thirty percent of the ionizing photons in the
center of II~Zw~40 are visible in the optical. The dust extinction in
II~Zw~40 is consistent with the clumpy and porous dust seen in other
starburst galaxies, regardless of their metallicity. The measured
extinction values appear to be luminosity weighted and thus biased
towards regions with little dust. 

Although disentangling the relative effects of mass and metallicity is
difficult, the star formation properties of II~Zw~40 described above
appear to be the result of the starburst ignited by its recent merger;
the low mass and low metallicity of this galaxy only appear to play a
secondary role. II~Zw~40's dust extinction is more similar to the
values seen in higher mass, higher metallicity starburst galaxies than
to the values seen in other low-metallicity dwarf galaxies. Its \hii\
region luminosity function is relatively flat, which is similar to
that seen in irregular galaxies. However, this is likely the effect of
the high ISM pressures created in merger-related starburst rather than
galaxy mass or metallicity.

Merger-driven bursts of star formation similar to the burst in
II~Zw~40 may be more common at high redshifts. Evidence from optical
studies suggests that massive galaxies at high redshift have
morphologies similar to today's low-mass irregular galaxies rather
than to today's spirals \citep{2009ApJ...701..306E}. Dwarf galaxies
are also the most abundant type of galaxy in the universe
\citep{2012dgkg.book..175B}. While the merger rate is relatively low
today, higher merger rates in the past mean that star formation events
like the one in II~Zw~40 may have been more likely in the early
universe \citep{2011ApJ...742..103L}.

\acknowledgments

A.A.K.\ thanks Laura Chomiuk, Deidre Hunter, Remy Indebetouw, Suzanne
Madden, Elizabeth Stanway, and David Whelan for helpful discussions
and Adam Leroy for detailed comments on an earlier draft.  K.E.J.\
acknowledges support provided by NSF through CAREER award 0548103 and
the David and Lucile Packard Foundation through a Packard Fellowship.
A.E.R.\ gratefully acknowledges support provided by NASA through the
Einstein Fellowship Program grant PF1-120086 and from a NASA Earth and
Space Science Fellowship.


\begin{deluxetable}{cclllll}
\tablewidth{0pt}
\tabletypesize{\scriptsize}
\tablecaption{Summary of VLA observations \label{tab:vla_obs_summary}}
\tablehead{
\colhead{Wavelength} &
\colhead{Frequency} &
\colhead{Array} &
\colhead{Date} &
\colhead{Program} &
\colhead{Flux} & 
\colhead{Phase} \\
\colhead{cm} & 
\colhead{GHz} &
\colhead{Config.} & 
\colhead{Observed} & 
\colhead{Code} & 
\colhead{Calibrator} & 
\colhead{Calibrator} }
\startdata
6.2  & 4.86  &  A   &  1999 August 23             & AT227 & 1331+305 & 0532+075  \\
6.2  & 4.86  & A+PT &  2006 March 14, 20-21       & AJ324 & 1331+305 & 0552+032  \\
6.2  & 4.86  & A    &  2006 March 26-27           & AJ324 & 1331+305 & 0552+032  \\
3.5  & 8.46  &  A   &  1999 August 23             & AT227 & 1331+305 & 0532+075  \\
3.5  & 8.46  &  A   &  2002 February 15-16        & AJ286 & 1331+305 & 0552+032  \\ 
1.3  & 22.46 &  A   &  1999 August 23             & AT227 & 1331+305 & 0552+032  \\ 
1.3  & 22.46 & A+PT &  2002 January 30            & AT272 & 0542+498 & 0552+032  \\
1.3  & 22.46 &  B   &  2006 August 20, 25-28      & AJ324 & 1331+305 & 0552+032  \\
\enddata
\end{deluxetable}

\begin{deluxetable}{ccccccc}
\tablecolumns{7}
\tablewidth{0pt}
\tablecaption{Summary of Radio Continuum Image Properties \label{tab:image_summary}}
\tablehead{
\colhead{Wavelength} &
\colhead{Frequency} &
\colhead{} &
\colhead{UV range} &
\colhead{Beam} &
\colhead{PA} &
\colhead{Noise} \\
\colhead{cm} &
\colhead{GHz} & 
\colhead{Robust} &
\colhead{k$\lambda$} &
\colhead{\arcsec} &
\colhead{\degr} &
\colhead{\uJybeam} }
\startdata
\cutinhead{Matched {\em u-v} coverage images} 
6.2    & 4.86  & 0 & 11.2 -- 1000 & $0.41 \times 0.24$ & 21.6  & 10.7 \\
3.5    & 8.46  & 0 & 11.2 -- 1000 & $0.41 \times 0.24$ & 21.6  & 23.7 \\
1.3    & 22.46 & 0  & 11.2 -- 1000 & $0.41 \times 0.24$ & 21.6  & 38.3 \\
\cutinhead{High resolution image}
1.3    & 22.46 & 0  & 11.2 -- 5500 & $0.14 \times 0.12$ & 28.2   & 24.0 \\
\enddata
\end{deluxetable}

\begin{deluxetable}{cccc}
\tablewidth{0pt}
\tablecaption{{\em HST} Flux Densities for Optical Clusters \label{tab:optical_fluxes_hst}}
\tablecolumns{4}
\tablehead{
    \colhead{} &
    \colhead{$f_{F555W}$} &
    \colhead{$f_{F658N}$} &
    \colhead{$f_{F814W}$} \\ 
    \colhead{Name} &
    \colhead{$10^{-18} \ {\rm erg \, s^{-1} \, cm^{-2} \, \AA^{-1}} $} &
    \colhead{$10^{-18} \ {\rm erg \, s^{-1} \, cm^{-2} \, \AA^{-1}} $} &
    \colhead{$10^{-18} \ {\rm erg \, s^{-1} \, cm^{-2} \, \AA^{-1}} $} }
\startdata
SSC-N & $  163 $ & $  751  $ &$   65 $    \\ 
SSC-S & $   60 $ & $   87  $ &$   48 $    \\ 
\enddata
    \tablecomments{The errors on the flux densities in all filters is 10\%.}
\end{deluxetable}

\begin{deluxetable}{cccccc}
\tablewidth{0pt}
\tablecaption{Radio Continuum Spectrum of the Central Compact Region of II Zw 40 \label{tab:radio_sed}}
\tablecolumns{6}
\tablehead{
    \colhead{Wavelength} &
    \colhead{Frequency} &
    \colhead{Flux Density\tablenotemark{a}} &
    \colhead{Error\tablenotemark{b}} & 
    \colhead{Total Flux Density\tablenotemark{c}} &
    \colhead{Missing Flux} \\
    \colhead{cm} &
    \colhead{GHz} &
    \colhead{mJy} &
    \colhead{mJy} &
    \colhead{mJy} &
    \colhead{\%}}
\startdata
1.3 & 22.46 & 7.0 & 1.4 & 17.9 & -61.0 \\ 
3.5 &  8.46 & 8.0 & 1.6 & 19.6 & -58.9 \\ 
6.2 &  4.86 & 10.1 & 2.0 & 20.6 & -51.0 \\ 
\enddata
\tablenotetext{a}{Flux densities derived from matched {\em u-v} coverage images. See~\S~\ref{sec:vla-data} and \ref{sec:radio-spectrum-ii} for details.}
\tablenotetext{b}{The errors are estimated to be 20\% based on typical VLA absolute flux density calibration uncertainties. }
\tablenotetext{c}{Estimated using model from \citet{1991A&A...246..323K}}
\end{deluxetable}

\begin{deluxetable}{lccccc}
\tabletypesize{\scriptsize}
\tablewidth{0pt}
\tablecaption{1.3\,cm Radio Continuum Source Flux Densities \label{tab:source_fluxes}}
\tablecolumns{6}
\tablehead{
    \colhead{} & 
    \colhead{RA} &
    \colhead{Dec} &
    \colhead{Compact (C)} &
    \colhead{Peak Flux} &
    \colhead{Flux Density\tablenotemark{a}} \\
    \colhead{Name} &
    \colhead{J2000} &
    \colhead{J2000} &
    \colhead{or Diffuse (D)} &
    \colhead{\mJybeam} &
    \colhead{mJy} }
\startdata
  A &   5 55 42.589 &  3 23 31.86 & D & 0.18  & $ 0.08 \pm 0.04 $  \\ 
  B &   5 55 42.596 &  3 23 31.41 & D & 0.15  & $ 0.05 \pm 0.04 $  \\ 
  C &   5 55 42.603 &  3 23 31.73 & D & 0.17  & $ 0.04 \pm 0.04 $  \\ 
  D\tablenotemark{b} &   5 55 42.609 &  3 23 31.98 & C & 0.42  & $ 0.67 \pm 0.03 $  \\ 
  E &   5 55 42.614 &  3 23 32.35 & C & 0.15  & $ 0.06 \pm 0.05 $  \\ 
  F &   5 55 42.614 &  3 23 31.67 & D & 0.15  & $ 0.03 \pm 0.04 $  \\ 
  G\tablenotemark{b} &   5 55 42.616 &  3 23 32.02 & C & 0.39  & $ 0.36 \pm 0.03 $  \\ 
  H &   5 55 42.623 &  3 23 32.12 & C & 0.33  & $ 0.34 \pm 0.03 $  \\ 
  I &   5 55 42.626 &  3 23 32.26 & C & 0.15  & $ 0.04 \pm 0.06 $  \\ 
  J &   5 55 42.632 &  3 23 31.82 & D & 0.19  & $ 0.10 \pm 0.03 $  \\ 
  K &   5 55 42.642 &  3 23 32.05 & C & 0.27  & $ 0.16 \pm 0.04 $  \\ 
\cutinhead{Combined Regions\tablenotemark{c}} \ DG & \nodata & \nodata & \nodata & \nodata & $ 0.87 \pm 0.03 $  \\ 
DGH & \nodata & \nodata & \nodata & \nodata & $ 1.24 \pm 0.03 $  \\ 
Entire Region\tablenotemark{d} & \nodata & \nodata & \nodata & \nodata & $ 4.42 \pm 0.03 $  \\ 
\enddata 

\tablenotetext{a}{Since these sources are compared to each other and
not to measurements at other wavelengths, the errors on the flux
densities are the statistical errors and do not take into account the
20\% uncertainty in the  photometry. }
    \tablenotetext{b}{The individual fluxes for sources D and G are very uncertain because we barely resolve them as two separate sources.}
\tablenotetext{c}{The properties of the combined regions were determined using regions including all the named sources rather than simply adding together the values for the individual sources.}
\tablenotetext{d}{The flux density determined here is lower than the
  flux density determined in Table~\ref{tab:radio_sed} because here we are
  using an aperture defined by the 3$\sigma$ contour of the high
  resolution image and there we used an aperture corresponding to the
  3$\sigma$ contour in the lower resolution, matched beam image.}
\end{deluxetable}

\begin{deluxetable}{ccccc}
\tablewidth{0pt}
\tablecaption{Properties of the 1.3\,cm Radio Continuum Sources \label{tab:source_properties}}
\tablecolumns{5}
\tablehead{
    \colhead{} &
    \colhead{$N_{ion}$} &
    \colhead{}&
    \colhead{$M_{cluster}$} &
    \colhead{SFR} \\
    \colhead{Name} &
    \colhead{$10^{51} \ {\rm s^{-1}}$} &
    \colhead{$N_{ostars}$} &
    \colhead{$10^4$ \Msun} &
    \colhead{$M_\sun \, yr^{-1}$} }
\startdata
  A & 0.8 &    80 & 1.6 & \nodata \\ 
  B & 0.5 &    50 & 1.0 & \nodata \\ 
  C & 0.5 &    50 & 0.9 & \nodata \\ 
  D & 7.1 &   710 & 14.2 & \nodata \\ 
  E & 0.7 &    70 & 1.3 & \nodata \\ 
  F & 0.3 &    30 & 0.6 & \nodata \\ 
  G & 3.8 &   380 & 7.6 & \nodata \\ 
  H & 3.6 &   360 & 7.1 & \nodata \\ 
  I & 0.4 &    40 & 0.9 & \nodata \\ 
  J & 1.1 &   110 & 2.1 & \nodata \\ 
  K & 1.7 &   170 & 3.4 & \nodata \\ 
\cutinhead{Combined Regions\tablenotemark{a}} \ DG & 9.2 &   920 & \nodata &  0.07 \\ 
DGH & 13.0 &  1300 & \nodata &  0.09 \\ 
Entire Region & 46.6 &  4660 & \nodata &  0.34 \\ 
\enddata
\tablenotetext{a}{The properties of the combined regions were determined using regions including all the named sources rather than simply adding together the values for the individual sources.}
\end{deluxetable}

\begin{deluxetable}{ccccccc}
\tablewidth{0pt}
\tablecaption{Properties of the Optical Clusters \label{tab:source_properties_hst}}
\tablecolumns{7}
\tablehead{
    \colhead{} &
    \colhead{ \ha\ flux} &
    \colhead{ \ha\ flux (extinction corrected)\tablenotemark{a}} &
    \colhead{$N_{ion}$} &
    \colhead{}&
    \colhead{Age\tablenotemark{b}} & 
    \colhead{Mass\tablenotemark{c}} \\
    \colhead{Name} &
    \colhead{$10^{-15} \ {\rm erg \, s^{-1} \, cm^{-2}}$} &
    \colhead{$10^{-15} \ {\rm erg \, s^{-1} \, cm^{-2}}$} &
    \colhead{$10^{51} \ {\rm s^{-1}}$} &
    \colhead{$N_{Ostars}$} &
    \colhead{Myr} &
    \colhead{$10^4 \ {\rm M_\odot}$}}
\startdata
SSC-N & $ 50.1 \pm  6.1 $ & $ 182.0 \pm  22.2 $ &  $\gtrsim   2.03$ &  $ \gtrsim   200 $ & $ \lesssim 5.0 $ &  9.0 \\ 
SSC-S & $  2.5 \pm  1.1 $ & $   9.3 \pm   4.1 $ &  0.10 &    10 & $ 9.4 ^{+1.4}_{-1.0} $  & 16.0 \\ 
\enddata
    \tablenotetext{a}{Corrected for foreground Galactic extinction using an E(\bv)=0.57.}
    \tablenotetext{b}{Derived from \ha\ equivalent widths.}
    \tablenotetext{c}{Derived using broadband fluxes. See Table~\ref{tab:optical_fluxes_hst} for values.}
\end{deluxetable}

\begin{deluxetable}{lccc}
\tablewidth{0pt}
\tablecaption{{\em HST} Flux Densities for Radio Sources \label{tab:optical_fluxes}}
\tablecolumns{4}
\tablehead{
    \colhead{} &
    \colhead{$f_{F555W}$} &
    \colhead{$f_{F658N}$} &
    \colhead{$f_{F814W}$} \\ 
    \colhead{Name} &
    \colhead{$10^{-18} \ {\rm erg \, s^{-1} \, cm^{-2} \, \AA^{-1}} $} &
    \colhead{$10^{-18} \ {\rm erg \, s^{-1} \, cm^{-2} \, \AA^{-1}} $} &
    \colhead{$10^{-18} \ {\rm erg \, s^{-1} \, cm^{-2} \, \AA^{-1}} $} }
\startdata
  A & $  5.28 \pm  0.22 $ & $ 36.70 \pm  1.11 $ &$  1.36 \pm  0.06 $    \\ 
  B & $  2.15 \pm  0.13 $ & $ 13.40 \pm  1.17 $ &$  0.35 \pm  0.04 $    \\ 
  C & $  9.85 \pm  0.82 $ & $ 88.70 \pm  7.96 $ &$  2.25 \pm  0.19 $    \\ 
  D\tablenotemark{a} & $ 67.30 \pm  0.30 $ & $ 493.00 \pm  2.51 $ &$ 17.20 \pm  0.09 $    \\ 
  E & $  1.19 \pm  0.07 $ & $  7.52 \pm  0.79 $ &$  0.41 \pm  0.05 $    \\ 
  F & $  1.34 \pm  1.16 $ & $ 25.50 \pm  8.27 $ &$  1.79 \pm  0.23 $    \\ 
  G\tablenotemark{a} & $ 82.10 \pm  0.50 $ & $ 296.00 \pm  3.74 $ &$ 35.30 \pm  0.14 $    \\ 
  H & $ 40.20 \pm  0.24 $ & $ 246.00 \pm  1.85 $ &$ 14.00 \pm  0.08 $    \\ 
  I & $  1.99 \pm  0.17 $ & $ 11.70 \pm  2.50 $ &$  0.87 \pm  0.06 $    \\ 
  J & $ 17.50 \pm  0.24 $ & $ 118.00 \pm  2.13 $ &$  3.77 \pm  0.08 $    \\ 
  K & $ 15.00 \pm  0.38 $ & $ 133.00 \pm  5.00 $ &$  2.97 \pm  0.16 $    \\ 
\cutinhead{Combined Regions\tablenotemark{b}} \ DG & $ 116.00 \pm  0.41 $ & $ 599.00 \pm  3.21 $ &$ 43.10 \pm  0.14 $    \\ 
DGH & $ 151.00 \pm  0.33 $ & $ 849.00 \pm  2.16 $ &$ 56.00 \pm  0.13 $    \\ 
Entire Region & $ 678.00 \pm  0.64 $ & $ 4640.00 \pm  4.01 $ &$ 197.00 \pm  0.24 $    \\ 
\enddata
  \tablenotetext{a}{The individual fluxes for sources D and G are very uncertain because we barely resolve them as two separate sources.}
\tablenotetext{b}{The properties of the combined regions were determined using regions including all the named sources rather than simply adding together the values for the individual sources.}
\end{deluxetable}

\begin{deluxetable}{lccccc}
\tabletypesize{\scriptsize}
\tablewidth{0pt}
\tablecaption{Extinction Values Derived for Radio Continuum Sources \label{tab:extinction}}
\tablecolumns{6}
\tablehead{
    \colhead{} &
    \colhead{Est. F658N Continuum\tablenotemark{a}} &
    \colhead{$f_{H\alpha}$\tablenotemark{b}} &
    \colhead{$f_{H\alpha,c}$\tablenotemark{c}} &
    \colhead{$f_{H\alpha,est}$\tablenotemark{d}} &
    \colhead{$A_{H\alpha}$\tablenotemark{e}} \\
    \colhead{Name} &
    \colhead{$10^{-16} \ {\rm erg \, s^{-1} \, cm^{-2} \, \AA^{-1}} $} &
    \colhead{$10^{-14} \ {\rm erg \, s^{-1} \, cm^{-2}}$} &
    \colhead{$10^{-14} \ {\rm erg \, s^{-1} \, cm^{-2}}$} &
    \colhead{$10^{-14} \ {\rm erg \, s^{-1} \, cm^{-2}}$} &
    \colhead{mag} }
\startdata
  A & $ 0.035 \pm 0.009  $ &  $  0.26 \pm  0.01 $ & $  0.94 \pm  0.06 $ & $     7 \pm     4 $ & $  2.22 \pm  0.30 $ \\ 
  B & $ 0.014 \pm 0.005  $ &  $  0.09 \pm  0.01 $ & $  0.34 \pm  0.04 $ & $     5 \pm     4 $ & $  2.83 \pm  0.45 $ \\ 
  C & $ 0.065 \pm 0.032  $ &  $  0.64 \pm  0.07 $ & $  2.33 \pm  0.26 $ & $     4 \pm     4 $ & $  0.61 \pm  0.53 $ \\ 
  D & $ 0.450 \pm 0.012  $ &  $  3.49 \pm  0.02 $ & $ 12.70 \pm  0.55 $ & $    64 \pm     8 $ & $  1.75 \pm  0.08 $ \\ 
  E & $ 0.008 \pm 0.003  $ &  $  0.05 \pm  0.01 $ & $  0.19 \pm  0.03 $ & $     6 \pm     5 $ & $  3.76 \pm  0.47 $ \\ 
  F & $ 0.015 \pm 0.045  $ &  $  0.19 \pm  0.07 $ & $  0.68 \pm  0.27 $ & $     3 \pm     4 $ & $  1.57 \pm  0.87 $ \\ 
  G & $ 0.612 \pm 0.019  $ &  $  1.83 \pm  0.03 $ & $  6.66 \pm  0.31 $ & $    34 \pm     5 $ & $  1.77 \pm  0.09 $ \\ 
  H & $ 0.286 \pm 0.009  $ &  $  1.70 \pm  0.02 $ & $  6.16 \pm  0.27 $ & $    32 \pm     5 $ & $  1.79 \pm  0.10 $ \\ 
  I & $ 0.015 \pm 0.007  $ &  $  0.08 \pm  0.02 $ & $  0.29 \pm  0.07 $ & $     4 \pm     6 $ & $  2.82 \pm  0.87 $ \\ 
  J & $ 0.114 \pm 0.009  $ &  $  0.83 \pm  0.02 $ & $  3.01 \pm  0.15 $ & $    10 \pm     3 $ & $  1.26 \pm  0.19 $ \\ 
  K & $ 0.096 \pm 0.015  $ &  $  0.96 \pm  0.04 $ & $  3.48 \pm  0.21 $ & $    15 \pm     4 $ & $  1.60 \pm  0.16 $ \\ 
\cutinhead{Combined Regions} \ DG & $ 0.833 \pm 0.016  $ &  $  4.02 \pm  0.03 $ & $ 14.60 \pm  0.64 $ & $    82 \pm    10 $ & $  1.88 \pm  0.08 $ \\ 
DGH & $ 1.090 \pm 0.013  $ &  $  5.77 \pm  0.02 $ & $ 21.00 \pm  0.91 $ & $   117 \pm    15 $ & $  1.86 \pm  0.08 $ \\ 
Entire Region & $ 4.640 \pm 0.025  $ &  $ 32.60 \pm  0.04 $ & $ 118.00 \pm  5.11 $ & $   417 \pm    52 $ & $  1.37 \pm  0.08 $ \\ 
\enddata
    \tablenotetext{a}{Continuum flux density in the F658N filter estimated by interpolating between the individual source flux densities in the F555W and F814W filters. See Section~\ref{sec:prop-optic-clust} for further details.}
    \tablenotetext{b}{Estimated optical \ha\ flux determined by subtracting the estimated continuum flux density in the F658N filter from the total flux density in the F658N filter and then multiplying the result by the filter width (see Section~\ref{sec:prop-optic-clust} for further details).}
    \tablenotetext{c}{Estimated optical \ha\ flux corrected for foreground Galactic extinction as described in Section~\ref{sec:indiv-source-extinct}.}
    \tablenotetext{d}{\ha\ flux predicted from the radio data and Equation~\ref{eq:f_ha_est}.}
    \tablenotetext{e}{Extinction at the wavelength of \ha.}
\end{deluxetable}


\begin{thebibliography}{95}
\expandafter\ifx\csname natexlab\endcsname\relax\def\natexlab#1{#1}\fi

\bibitem[{{Anders} \& {Fritze-v.~Alvensleben}(2003)}]{2003A&A...401.1063A}
{Anders}, P., \& {Fritze-v.~Alvensleben}, U. 2003, \aap, 401, 1063

\bibitem[{{Baars} {et~al.}(1977){Baars}, {Genzel}, {Pauliny-Toth}, \&
  {Witzel}}]{1977A&A....61...99B}
{Baars}, J.~W.~M., {Genzel}, R., {Pauliny-Toth}, I.~I.~K., \& {Witzel}, A.
  1977, \aap, 61, 99

\bibitem[{{Bastian} {et~al.}(2010){Bastian}, {Covey}, \&
  {Meyer}}]{2010ARA&A..48..339B}
{Bastian}, N., {Covey}, K.~R., \& {Meyer}, M.~R. 2010, \araa, 48, 339

\bibitem[{{Beck} \& {Krause}(2005)}]{2005AN....326..414B}
{Beck}, R., \& {Krause}, M. 2005, Astronomische Nachrichten, 326, 414

\bibitem[{{Beck} {et~al.}(2002){Beck}, {Turner}, {Langland-Shula}, {Meier},
  {Crosthwaite}, \& {Gorjian}}]{2002AJ....124.2516B}
{Beck}, S.~C., {Turner}, J.~L., {Langland-Shula}, L.~E., {Meier}, D.~S.,
  {Crosthwaite}, L.~P., \& {Gorjian}, V. 2002, \aj, 124, 2516

\bibitem[{{Bell}(2003)}]{2003ApJ...586..794B}
{Bell}, E.~F. 2003, \apj, 586, 794

\bibitem[{{Bergvall}(2012)}]{2012dgkg.book..175B}
{Bergvall}, N. 2012, {Star Forming Dwarf Galaxies}, ed. P.~{Papaderos},
  S.~{Recchi}, \& G.~{Hensler}, 175

\bibitem[{{Bigiel} {et~al.}(2008){Bigiel}, {Leroy}, {Walter}, {Brinks}, {de
  Blok}, {Madore}, \& {Thornley}}]{2008AJ....136.2846B}
{Bigiel}, F., {Leroy}, A., {Walter}, F., {Brinks}, E., {de Blok}, W.~J.~G.,
  {Madore}, B., \& {Thornley}, M.~D. 2008, \aj, 136, 2846

\bibitem[{{Bolatto} {et~al.}(2008){Bolatto}, {Leroy}, {Rosolowsky}, {Walter},
  \& {Blitz}}]{2008ApJ...686..948B}
{Bolatto}, A.~D., {Leroy}, A.~K., {Rosolowsky}, E., {Walter}, F., \& {Blitz},
  L. 2008, \apj, 686, 948

\bibitem[{{Brinks} \& {Klein}(1988)}]{1988MNRAS.231P..63B}
{Brinks}, E., \& {Klein}, U. 1988, \mnras, 231, 63P

\bibitem[{{Burstein} \& {Heiles}(1982)}]{1982AJ.....87.1165B}
{Burstein}, D., \& {Heiles}, C. 1982, \aj, 87, 1165

\bibitem[{{Calzetti}(2001)}]{2001PASP..113.1449C}
{Calzetti}, D. 2001, \pasp, 113, 1449

\bibitem[{{Calzetti} {et~al.}(2000){Calzetti}, {Armus}, {Bohlin}, {Kinney},
  {Koornneef}, \& {Storchi-Bergmann}}]{2000ApJ...533..682C}
{Calzetti}, D., {Armus}, L., {Bohlin}, R.~C., {Kinney}, A.~L., {Koornneef}, J.,
  \& {Storchi-Bergmann}, T. 2000, \apj, 533, 682

\bibitem[{{Calzetti} {et~al.}(2007){Calzetti}, {Kennicutt}, {Engelbracht},
  {Leitherer}, {Draine}, {Kewley}, {Moustakas}, {Sosey}, {Dale}, {Gordon},
  {Helou}, {Hollenbach}, {Armus}, {Bendo}, {Bot}, {Buckalew}, {Jarrett}, {Li},
  {Meyer}, {Murphy}, {Prescott}, {Regan}, {Rieke}, {Roussel}, {Sheth}, {Smith},
  {Thornley}, \& {Walter}}]{2007ApJ...666..870C}
{Calzetti}, D., {Kennicutt}, R.~C., {Engelbracht}, C.~W., {Leitherer}, C.,
  {Draine}, B.~T., {Kewley}, L., {Moustakas}, J., {Sosey}, M., {Dale}, D.~A.,
  {Gordon}, K.~D., {Helou}, G.~X., {Hollenbach}, D.~J., {Armus}, L., {Bendo},
  G., {Bot}, C., {Buckalew}, B., {Jarrett}, T., {Li}, A., {Meyer}, M.,
  {Murphy}, E.~J., {Prescott}, M., {Regan}, M.~W., {Rieke}, G.~H., {Roussel},
  H., {Sheth}, K., {Smith}, J.~D.~T., {Thornley}, M.~D., \& {Walter}, F. 2007,
  \apj, 666, 870

\bibitem[{{Cardelli} {et~al.}(1989){Cardelli}, {Clayton}, \&
  {Mathis}}]{1989ApJ...345..245C}
{Cardelli}, J.~A., {Clayton}, G.~C., \& {Mathis}, J.~S. 1989, \apj, 345, 245

\bibitem[{{Chomiuk} \& {Povich}(2011)}]{2011AJ....142..197C}
{Chomiuk}, L., \& {Povich}, M.~S. 2011, \aj, 142, 197

\bibitem[{{Chomiuk} \& {Wilcots}(2009{\natexlab{a}})}]{2009AJ....137.3869C}
{Chomiuk}, L., \& {Wilcots}, E.~M. 2009{\natexlab{a}}, \aj, 137, 3869

\bibitem[{{Chomiuk} \& {Wilcots}(2009{\natexlab{b}})}]{2009ApJ...703..370C}
---. 2009{\natexlab{b}}, \apj, 703, 370

\bibitem[{{Condon}(1992)}]{1992ARA&A..30..575C}
{Condon}, J.~J. 1992, \araa, 30, 575

\bibitem[{{Davies} {et~al.}(1998){Davies}, {Sugai}, \&
  {Ward}}]{1998MNRAS.295...43D}
{Davies}, R.~I., {Sugai}, H., \& {Ward}, M.~J. 1998, \mnras, 295, 43

\bibitem[{{Deeg} {et~al.}(1993){Deeg}, {Brinks}, {Duric}, {Klein}, \&
  {Skillman}}]{1993ApJ...410..626D}
{Deeg}, H., {Brinks}, E., {Duric}, N., {Klein}, U., \& {Skillman}, E. 1993,
  \apj, 410, 626

\bibitem[{{Dufour}(1984)}]{1984IAUS..108..353D}
{Dufour}, R.~J. 1984, in IAU Symposium, Vol. 108, Structure and Evolution of
  the Magellanic Clouds, ed. S.~{van den Bergh} \& K.~S.~D. {de Boer}, 353--360

\bibitem[{{Duric} {et~al.}(1988){Duric}, {Bourneuf}, \&
  {Gregory}}]{1988AJ.....96...81D}
{Duric}, N., {Bourneuf}, E., \& {Gregory}, P.~C. 1988, \aj, 96, 81

\bibitem[{{Elmegreen} \& {Efremov}(1997)}]{1997ApJ...480..235E}
{Elmegreen}, B.~G., \& {Efremov}, Y.~N. 1997, \apj, 480, 235

\bibitem[{{Elmegreen} {et~al.}(2009){Elmegreen}, {Elmegreen}, {Marcus},
  {Shahinyan}, {Yau}, \& {Petersen}}]{2009ApJ...701..306E}
{Elmegreen}, D.~M., {Elmegreen}, B.~G., {Marcus}, M.~T., {Shahinyan}, K.,
  {Yau}, A., \& {Petersen}, M. 2009, \apj, 701, 306

\bibitem[{{Fall} {et~al.}(2009){Fall}, {Chandar}, \&
  {Whitmore}}]{2009ApJ...704..453F}
{Fall}, S.~M., {Chandar}, R., \& {Whitmore}, B.~C. 2009, \apj, 704, 453

\bibitem[{{Galliano} {et~al.}(2005){Galliano}, {Madden}, {Jones}, {Wilson}, \&
  {Bernard}}]{2005A&A...434..867G}
{Galliano}, F., {Madden}, S.~C., {Jones}, A.~P., {Wilson}, C.~D., \& {Bernard},
  J. 2005, \aap, 434, 867

\bibitem[{{Galliano} {et~al.}(2003){Galliano}, {Madden}, {Jones}, {Wilson},
  {Bernard}, \& {Le Peintre}}]{2003A&A...407..159G}
{Galliano}, F., {Madden}, S.~C., {Jones}, A.~P., {Wilson}, C.~D., {Bernard},
  J., \& {Le Peintre}, F. 2003, \aap, 407, 159

\bibitem[{{Gil de Paz} {et~al.}(2003){Gil de Paz}, {Madore}, \&
  {Pevunova}}]{2003ApJS..147...29G}
{Gil de Paz}, A., {Madore}, B.~F., \& {Pevunova}, O. 2003, \apjs, 147, 29

\bibitem[{{Gordon} {et~al.}(1997){Gordon}, {Calzetti}, \&
  {Witt}}]{1997ApJ...487..625G}
{Gordon}, K.~D., {Calzetti}, D., \& {Witt}, A.~N. 1997, \apj, 487, 625

\bibitem[{Greisen(2010)}]{ch2:aipscookbook}
Greisen, E., ed. 2010, AIPS COOKBOOK (NRAO)

\bibitem[{{Guseva} {et~al.}(2000){Guseva}, {Izotov}, \&
  {Thuan}}]{2000ApJ...531..776G}
{Guseva}, N.~G., {Izotov}, Y.~I., \& {Thuan}, T.~X. 2000, \apj, 531, 776

\bibitem[{{Guti{\'e}rrez} {et~al.}(2011){Guti{\'e}rrez}, {Beckman}, \&
  {Buenrostro}}]{2011AJ....141..113G}
{Guti{\'e}rrez}, L., {Beckman}, J.~E., \& {Buenrostro}, V. 2011, \aj, 141, 113

\bibitem[{{Hunt} {et~al.}(2005){Hunt}, {Bianchi}, \&
  {Maiolino}}]{2005A&A...434..849H}
{Hunt}, L., {Bianchi}, S., \& {Maiolino}, R. 2005, \aap, 434, 849

\bibitem[{{Hunt} {et~al.}(2004){Hunt}, {Dyer}, {Thuan}, \&
  {Ulvestad}}]{2004ApJ...606..853H}
{Hunt}, L.~K., {Dyer}, K.~K., {Thuan}, T.~X., \& {Ulvestad}, J.~S. 2004, \apj,
  606, 853

\bibitem[{{Hunter} \& {Gallagher}(1997)}]{1997ApJ...475...65H}
{Hunter}, D.~A., \& {Gallagher}, III, J.~S. 1997, \apj, 475, 65

\bibitem[{{Inoue}(2001)}]{2001AJ....122.1788I}
{Inoue}, A.~K. 2001, \aj, 122, 1788

\bibitem[{{Izotov} \& {Thuan}(2011)}]{2011ApJ...734...82I}
{Izotov}, Y.~I., \& {Thuan}, T.~X. 2011, \apj, 734, 82

\bibitem[{{Johnson} {et~al.}(2009){Johnson}, {Hunt}, \&
  {Reines}}]{2009AJ....137.3788J}
{Johnson}, K.~E., {Hunt}, L.~K., \& {Reines}, A.~E. 2009, \aj, 137, 3788

\bibitem[{{Johnson} {et~al.}(2004){Johnson}, {Indebetouw}, {Watson}, \&
  {Kobulnicky}}]{2004AJ....128..610J}
{Johnson}, K.~E., {Indebetouw}, R., {Watson}, C., \& {Kobulnicky}, H.~A. 2004,
  \aj, 128, 610

\bibitem[{{Johnson} \& {Kobulnicky}(2003)}]{2003ApJ...597..923J}
{Johnson}, K.~E., \& {Kobulnicky}, H.~A. 2003, \apj, 597, 923

\bibitem[{{Joy} \& {Lester}(1988)}]{1988ApJ...331..145J}
{Joy}, M., \& {Lester}, D.~F. 1988, \apj, 331, 145

\bibitem[{{Kennicutt} \& {Evans}(2012)}]{2012ARA&A..50..531K}
{Kennicutt}, R.~C., \& {Evans}, N.~J. 2012, \araa, 50, 531

\bibitem[{{Kennicutt} {et~al.}(1989){Kennicutt}, {Edgar}, \&
  {Hodge}}]{1989ApJ...337..761K}
{Kennicutt}, Jr., R.~C., {Edgar}, B.~K., \& {Hodge}, P.~W. 1989, \apj, 337, 761

\bibitem[{{Klein} {et~al.}(1991){Klein}, {Weiland}, \&
  {Brinks}}]{1991A&A...246..323K}
{Klein}, U., {Weiland}, H., \& {Brinks}, E. 1991, \aap, 246, 323

\bibitem[{{Klein} {et~al.}(1984){Klein}, {Wielebinski}, \&
  {Thuan}}]{1984A&A...141..241K}
{Klein}, U., {Wielebinski}, R., \& {Thuan}, T.~X. 1984, \aap, 141, 241

\bibitem[{{Kobulnicky} \& {Johnson}(1999)}]{1999ApJ...527..154K}
{Kobulnicky}, H.~A., \& {Johnson}, K.~E. 1999, \apj, 527, 154

\bibitem[{{Kotulla} {et~al.}(2009){Kotulla}, {Fritze}, {Weilbacher}, \&
  {Anders}}]{2009MNRAS.396..462K}
{Kotulla}, R., {Fritze}, U., {Weilbacher}, P., \& {Anders}, P. 2009, \mnras,
  396, 462

\bibitem[{{Krumholz} {et~al.}(2010){Krumholz}, {Cunningham}, {Klein}, \&
  {McKee}}]{2010ApJ...713.1120K}
{Krumholz}, M.~R., {Cunningham}, A.~J., {Klein}, R.~I., \& {McKee}, C.~F. 2010,
  \apj, 713, 1120

\bibitem[{{Lacki} {et~al.}(2010){Lacki}, {Thompson}, \&
  {Quataert}}]{2010ApJ...717....1L}
{Lacki}, B.~C., {Thompson}, T.~A., \& {Quataert}, E. 2010, \apj, 717, 1

\bibitem[{{Leitherer}(1990)}]{1990ApJS...73....1L}
{Leitherer}, C. 1990, \apjs, 73, 1

\bibitem[{{Leitherer} {et~al.}(1999){Leitherer}, {Schaerer}, {Goldader},
  {Gonz{\'a}lez Delgado}, {Robert}, {Kune}, {de Mello}, {Devost}, \&
  {Heckman}}]{1999ApJS..123....3L}
{Leitherer}, C., {Schaerer}, D., {Goldader}, J.~D., {Gonz{\'a}lez Delgado},
  R.~M., {Robert}, C., {Kune}, D.~F., {de Mello}, D.~F., {Devost}, D., \&
  {Heckman}, T.~M. 1999, \apjs, 123, 3

\bibitem[{{Leroy} {et~al.}(2012){Leroy}, {Bigiel}, {de Blok}, {Boissier},
  {Bolatto}, {Brinks}, {Madore}, {Munoz-Mateos}, {Murphy}, {Sandstrom},
  {Schruba}, \& {Walter}}]{2012AJ....144....3L}
{Leroy}, A.~K., {Bigiel}, F., {de Blok}, W.~J.~G., {Boissier}, S., {Bolatto},
  A., {Brinks}, E., {Madore}, B., {Munoz-Mateos}, J.-C., {Murphy}, E.,
  {Sandstrom}, K., {Schruba}, A., \& {Walter}, F. 2012, \aj, 144, 3

\bibitem[{{Lotz} {et~al.}(2011){Lotz}, {Jonsson}, {Cox}, {Croton}, {Primack},
  {Somerville}, \& {Stewart}}]{2011ApJ...742..103L}
{Lotz}, J.~M., {Jonsson}, P., {Cox}, T.~J., {Croton}, D., {Primack}, J.~R.,
  {Somerville}, R.~S., \& {Stewart}, K. 2011, \apj, 742, 103

\bibitem[{{Madden} {et~al.}(2006){Madden}, {Galliano}, {Jones}, \&
  {Sauvage}}]{2006A&A...446..877M}
{Madden}, S.~C., {Galliano}, F., {Jones}, A.~P., \& {Sauvage}, M. 2006, \aap,
  446, 877

\bibitem[{{Malumuth} \& {Heap}(1994)}]{1994AJ....107.1054M}
{Malumuth}, E.~M., \& {Heap}, S.~R. 1994, \aj, 107, 1054

\bibitem[{{Meurer} {et~al.}(1997){Meurer}, {Heckman}, {Lehnert}, {Leitherer},
  \& {Lowenthal}}]{1997AJ....114...54M}
{Meurer}, G.~R., {Heckman}, T.~M., {Lehnert}, M.~D., {Leitherer}, C., \&
  {Lowenthal}, J. 1997, \aj, 114, 54

\bibitem[{{Mezger} {et~al.}(1967){Mezger}, {Schraml}, \&
  {Terzian}}]{1967ApJ...150..807M}
{Mezger}, P.~G., {Schraml}, J., \& {Terzian}, Y. 1967, \apj, 150, 807

\bibitem[{{Mills} {et~al.}(1978){Mills}, {Turtle}, \&
  {Watkinson}}]{1978MNRAS.185..263M}
{Mills}, B.~Y., {Turtle}, A.~J., \& {Watkinson}, A. 1978, \mnras, 185, 263

\bibitem[{{Murphy} {et~al.}(2011){Murphy}, {Condon}, {Schinnerer}, {Kennicutt},
  {Calzetti}, {Armus}, {Helou}, {Turner}, {Aniano}, {Beir{\~a}o}, {Bolatto},
  {Brandl}, {Croxall}, {Dale}, {Donovan Meyer}, {Draine}, {Engelbracht},
  {Hunt}, {Hao}, {Koda}, {Roussel}, {Skibba}, \& {Smith}}]{2011ApJ...737...67M}
{Murphy}, E.~J., {Condon}, J.~J., {Schinnerer}, E., {Kennicutt}, R.~C.,
  {Calzetti}, D., {Armus}, L., {Helou}, G., {Turner}, J.~L., {Aniano}, G.,
  {Beir{\~a}o}, P., {Bolatto}, A.~D., {Brandl}, B.~R., {Croxall}, K.~V.,
  {Dale}, D.~A., {Donovan Meyer}, J.~L., {Draine}, B.~T., {Engelbracht}, C.,
  {Hunt}, L.~K., {Hao}, C.-N., {Koda}, J., {Roussel}, H., {Skibba}, R., \&
  {Smith}, J.-D.~T. 2011, \apj, 737, 67

\bibitem[{{Murray} {et~al.}(2010){Murray}, {Quataert}, \&
  {Thompson}}]{2010ApJ...709..191M}
{Murray}, N., {Quataert}, E., \& {Thompson}, T.~A. 2010, \apj, 709, 191

\bibitem[{{Niklas} {et~al.}(1997){Niklas}, {Klein}, \&
  {Wielebinski}}]{1997A&A...322...19N}
{Niklas}, S., {Klein}, U., \& {Wielebinski}, R. 1997, \aap, 322, 19

\bibitem[{{Oey} \& {Clarke}(1998)}]{1998AJ....115.1543O}
{Oey}, M.~S., \& {Clarke}, C.~J. 1998, \aj, 115, 1543

\bibitem[{{Ostriker} \& {Shetty}(2011)}]{2011ApJ...731...41O}
{Ostriker}, E.~C., \& {Shetty}, R. 2011, \apj, 731, 41

\bibitem[{{Perley} \& {Butler}(2012)}]{2012arXiv1211.1300P}
{Perley}, R.~A., \& {Butler}, B.~J. 2012, ArXiv e-prints

\bibitem[{{Plante} \& {Sauvage}(2002)}]{2002AJ....124.1995P}
{Plante}, S., \& {Sauvage}, M. 2002, \aj, 124, 1995

\bibitem[{{Pleuss} {et~al.}(2000){Pleuss}, {Heller}, \&
  {Fricke}}]{2000A&A...361..913P}
{Pleuss}, P.~O., {Heller}, C.~H., \& {Fricke}, K.~J. 2000, \aap, 361, 913

\bibitem[{{Reines} \& {Deller}(2012)}]{2012ApJ...750L..24R}
{Reines}, A.~E., \& {Deller}, A.~T. 2012, \apjl, 750, L24

\bibitem[{{Reines} {et~al.}(2008{\natexlab{a}}){Reines}, {Johnson}, \&
  {Goss}}]{2008AJ....135.2222R}
{Reines}, A.~E., {Johnson}, K.~E., \& {Goss}, W.~M. 2008{\natexlab{a}}, \aj,
  135, 2222

\bibitem[{{Reines} {et~al.}(2008{\natexlab{b}}){Reines}, {Johnson}, \&
  {Hunt}}]{2008AJ....136.1415R}
{Reines}, A.~E., {Johnson}, K.~E., \& {Hunt}, L.~K. 2008{\natexlab{b}}, \aj,
  136, 1415

\bibitem[{{Reines} {et~al.}(2010){Reines}, {Nidever}, {Whelan}, \&
  {Johnson}}]{2010ApJ...708...26R}
{Reines}, A.~E., {Nidever}, D.~L., {Whelan}, D.~G., \& {Johnson}, K.~E. 2010,
  \apj, 708, 26

\bibitem[{{Reines} {et~al.}(2011){Reines}, {Sivakoff}, {Johnson}, \&
  {Brogan}}]{2011Natur.470...66R}
{Reines}, A.~E., {Sivakoff}, G.~R., {Johnson}, K.~E., \& {Brogan}, C.~L. 2011,
  \nat, 470, 66

\bibitem[{{Rela{\~n}o} {et~al.}(2012){Rela{\~n}o}, {Kennicutt}, {Eldridge},
  {Lee}, \& {Verley}}]{2012MNRAS.423.2933R}
{Rela{\~n}o}, M., {Kennicutt}, Jr., R.~C., {Eldridge}, J.~J., {Lee}, J.~C., \&
  {Verley}, S. 2012, \mnras, 423, 2933

\bibitem[{{Sargent} \& {Searle}(1970)}]{1970ApJ...162L.155S}
{Sargent}, W.~L.~W., \& {Searle}, L. 1970, \apjl, 162, L155+

\bibitem[{{Schlafly} \& {Finkbeiner}(2011)}]{2011ApJ...737..103S}
{Schlafly}, E.~F., \& {Finkbeiner}, D.~P. 2011, \apj, 737, 103

\bibitem[{{Schlegel} {et~al.}(1998){Schlegel}, {Finkbeiner}, \&
  {Davis}}]{1998ApJ...500..525S}
{Schlegel}, D.~J., {Finkbeiner}, D.~P., \& {Davis}, M. 1998, \apj, 500, 525

\bibitem[{{Scoville} {et~al.}(2001){Scoville}, {Polletta}, {Ewald}, {Stolovy},
  {Thompson}, \& {Rieke}}]{2001AJ....122.3017S}
{Scoville}, N.~Z., {Polletta}, M., {Ewald}, S., {Stolovy}, S.~R., {Thompson},
  R., \& {Rieke}, M. 2001, \aj, 122, 3017

\bibitem[{{Searle} \& {Sargent}(1972)}]{1972ApJ...173...25S}
{Searle}, L., \& {Sargent}, W.~L.~W. 1972, \apj, 173, 25

\bibitem[{{Sramek} \& {Weedman}(1986)}]{1986ApJ...302..640S}
{Sramek}, R.~A., \& {Weedman}, D.~W. 1986, \apj, 302, 640

\bibitem[{{Storey} \& {Hummer}(1995)}]{1995MNRAS.272...41S}
{Storey}, P.~J., \& {Hummer}, D.~G. 1995, \mnras, 272, 41

\bibitem[{{Thompson} {et~al.}(2005){Thompson}, {Quataert}, \&
  {Murray}}]{2005ApJ...630..167T}
{Thompson}, T.~A., {Quataert}, E., \& {Murray}, N. 2005, \apj, 630, 167

\bibitem[{{Thuan} {et~al.}(1999){Thuan}, {Sauvage}, \&
  {Madden}}]{1999ApJ...516..783T}
{Thuan}, T.~X., {Sauvage}, M., \& {Madden}, S. 1999, \apj, 516, 783

\bibitem[{{Tully} \& {Fisher}(1988)}]{1988cng..book.....T}
{Tully}, R.~B., \& {Fisher}, J.~R. 1988, {Catalog of Nearby Galaxies}, ed.
  {Tully, R.~B.~\& Fisher, J.~R.}

\bibitem[{{Ulvestad} {et~al.}(1998){Ulvestad}, {Beresford}, {Sowinski}, \&
  {Broadwell}}]{VLAtestmemo217}
{Ulvestad}, J., {Beresford}, R., {Sowinski}, K., \& {Broadwell}, C. 1998,
  {VLA-Pie Town Link Test Report}, VLA Test Memo 217, NRAO

\bibitem[{{Vacca}(1994)}]{1994ApJ...421..140V}
{Vacca}, W.~D. 1994, \apj, 421, 140

\bibitem[{{Vacca} {et~al.}(1996){Vacca}, {Garmany}, \&
  {Shull}}]{1996ApJ...460..914V}
{Vacca}, W.~D., {Garmany}, C.~D., \& {Shull}, J.~M. 1996, \apj, 460, 914

\bibitem[{{van Zee} {et~al.}(1998){van Zee}, {Skillman}, \&
  {Salzer}}]{1998AJ....116.1186V}
{van Zee}, L., {Skillman}, E.~D., \& {Salzer}, J.~J. 1998, \aj, 116, 1186

\bibitem[{{Vanzi} {et~al.}(2008){Vanzi}, {Cresci}, {Telles}, \&
  {Melnick}}]{2008A&A...486..393V}
{Vanzi}, L., {Cresci}, G., {Telles}, E., \& {Melnick}, J. 2008, \aap, 486, 393

\bibitem[{{Vanzi} {et~al.}(1996){Vanzi}, {Rieke}, {Martin}, \&
  {Shields}}]{1996ApJ...466..150V}
{Vanzi}, L., {Rieke}, G.~H., {Martin}, C.~L., \& {Shields}, J.~C. 1996, \apj,
  466, 150

\bibitem[{{Walborn}(1991)}]{1991IAUS..148..145W}
{Walborn}, N.~R. 1991, in IAU Symposium, Vol. 148, The Magellanic Clouds, ed.
  R.~{Haynes} \& D.~{Milne}, 145

\bibitem[{{Walsh} \& {Roy}(1993)}]{1993MNRAS.262...27W}
{Walsh}, J.~R., \& {Roy}, J. 1993, \mnras, 262, 27

\bibitem[{{Wang} \& {Heckman}(1996)}]{1996ApJ...457..645W}
{Wang}, B., \& {Heckman}, T.~M. 1996, \apj, 457, 645

\bibitem[{{Weidner} {et~al.}(2010){Weidner}, {Bonnell}, \&
  {Zinnecker}}]{2010ApJ...724.1503W}
{Weidner}, C., {Bonnell}, I.~A., \& {Zinnecker}, H. 2010, \apj, 724, 1503

\bibitem[{{Wynn-Williams} \& {Becklin}(1986)}]{1986ApJ...308..620W}
{Wynn-Williams}, C.~G., \& {Becklin}, E.~E. 1986, \apj, 308, 620

\bibitem[{{Youngblood} \& {Hunter}(1999)}]{1999ApJ...519...55Y}
{Youngblood}, A.~J., \& {Hunter}, D.~A. 1999, \apj, 519, 55

\end{thebibliography}
\end{document}